\documentclass[letter]{emulateapj}

\usepackage{natbib}
\usepackage{amssymb}
\usepackage{amsmath}
\usepackage{color}

\definecolor{myred}{rgb}{0.8,0.0,0.0}

\newcommand{\uKamin}{$\mu$K$_{\textrm{CMB}}$-amin}

\begin{document}

\title{Sunyaev-Zel'dovich-Measured Pressure Profiles 
  from the Bolocam X-ray/SZ Galaxy Cluster Sample}

\author{J.~Sayers\altaffilmark{1,9},
   N.~G.~Czakon\altaffilmark{1},
   A.~Mantz\altaffilmark{2},
   S.~R.~Golwala\altaffilmark{1},
   S.~Ameglio\altaffilmark{3},
   T.~P.~Downes\altaffilmark{1},
   P.~M.~Koch\altaffilmark{4},
   K.-Y.~Lin\altaffilmark{4},
   B.~J.~Maughan\altaffilmark{5},
   S.~M.~Molnar\altaffilmark{6},
   L.~Moustakas\altaffilmark{7},
   T.~Mroczkowski\altaffilmark{1,7,8},
   E.~Pierpaoli\altaffilmark{3},
   J.~A.~Shitanishi\altaffilmark{3},
   S.~Siegel\altaffilmark{1},
   K.~Umetsu\altaffilmark{4},
   and N.~Van~der~Pyl\altaffilmark{5}
 }
\altaffiltext{1}
  {Division of Physics, Math, and Astronomy, California Institute of Technology, Pasadena, CA 91125}
\altaffiltext{2}
  {Kavli Institute for Cosmological Physics, University of Chicago, 5640 South Ellis Avenue, Chicago, IL 60637}
\altaffiltext{3}
  {University of Southern California, Los Angeles, CA 90089}
\altaffiltext{4}
  {Institute of Astronomy and Astrophysics, Academia Sinica, 
    P.O. Box 23-141, Taipei 10617, Taiwan}
\altaffiltext{5}
  {H. H. Wills Physics Laboratory, University of Bristol, Tyndall Avenue, Bristol Bs8 ITL}
\altaffiltext{6}
  {LeCosPA Center, National Taiwan University, 
    Taipei 10617, Taiwan}
\altaffiltext{7}
  {Jet Propulsion Laboratory, Pasadena, CA 91109}
\altaffiltext{8}
  {NASA Einstein Postdoctoral Fellow}
\altaffiltext{9}
  {jack@caltech.edu}

\begin{abstract}

We describe Sunyaev-Zel'dovich (SZ) effect measurements and 
analysis of the intracluster medium (ICM) pressure
profiles of a set of 45 massive galaxy clusters imaged 
using Bolocam at the Caltech Submillimeter Observatory.
We deproject the average 
pressure profile of our sample into 13 logarithmically
spaced radial bins between 0.07$R_{500}$ and 3.5$R_{500}$, and
we find that a generalized Navarro, Frenk, and White (gNFW) profile
describes our data with sufficient goodness-of-fit and best-fit parameters
($C_{500}$, $\alpha$, $\beta$, $\gamma$, $P_0$ = 
1.18, 0.86, 3.67, 0.67, 4.29).
We use X-ray data to define
cool-core and disturbed subsamples of clusters,
and we constrain the average pressure profiles of each of these subsamples.
We find that, given the precision of our data, 
the average pressure profiles of disturbed and cool-core clusters
are consistent with one another at $R \gtrsim 0.15R_{500}$, with cool-core systems 
showing indications of
higher pressure at $R \lesssim 0.15R_{500}$. 
In addition, for the first time, we place simultaneous
constraints on the mass scaling of cluster pressure profiles,
their ensemble mean profile, and their radius-dependent intrinsic scatter
between 0.1$R_{500}$ and 2.0$R_{500}$.
The scatter among 
profiles is minimized at radii between $\simeq 0.2R_{500}$ and $\simeq 0.5R_{500}$,
with a value of $\simeq 20$\%.
These results for the intrinsic scatter
are largely consistent with previous analyses,
most of which have relied heavily on X-ray derived pressures of clusters
at significantly lower masses and redshifts compared to our sample.
Therefore, our data provide further evidence that cluster pressure
profiles are largely universal with scatter 
of $\simeq 20$--$40$\% about the universal profile over a wide
range of masses and redshifts.

\end{abstract}

\section{Introduction}

Massive galaxy clusters, the largest virialized systems in the universe,
appear to be exceptionally regular objects.
This is especially true at intermediate radii outside of the core 
(where complicated baryonic physics plays a large role),
and inside the actively accreting outer regions
(where non-equilibrium effects become significant).
In these intermediate regions simple, self-similar
scalings based on hydrostatic equilibrium and 
gravitational physics describe observations and simulations
quite well \citep{kaiser86, kravtsov12}.
Specifically, a range of observational results show approximately
universal behavior among mass and intracluster medium (ICM)
profiles after scaling by
characteristic overdensity radii such as $R_{500}$\footnote{
  $R_{500}$ denotes the radius where the average enclosed mass
  density is 500 times the critical density. Throughout this manuscript
  we define quantities at, or enclosed within, this characteristic radius.}
and by self-similar mass- and redshift-dependent normalizations 
(\citealt{vikhlinin06, pratt07, nagai07, cavagnolo09,arnaud10}, hereafter A10,
\citealt{planck12_v}, hereafter P12, \citealt{navarro96, umetsu11, kravtsov12, walker12}).
For example, the cluster-to-cluster dispersion (intrinsic scatter)
observed in these scaled profiles outside of the core regions
and inside $R_{500}$ is generally $\simeq 10$-- $40$\%
for entropy \citep{pratt06, cavagnolo09, pratt10, walker12},
gas density \citep{vikhlinin06, croston08, maughan12, eckert12},
temperature \citep{vikhlinin06, pratt07, leccardi08}, 
and pressure \citep[A10, ][]{sun11}.
In particular, both simulations and observations 
indicate low cluster-to-cluster dispersion
in pressure profiles at intermediate radii
(\citealt{borgani04, nagai07, piffaretti08}, A10, 
\citealt{plagge10, bonamente11}, P12).

Historically, observational studies of the ICM pressure have
relied almost exclusively on X-ray data.
These data have provided precise constraints on the
pressure profiles in the inner regions of clusters 
($R \le R_{500}$),
but the density-squared dependence of the X-ray surface
brightness makes it difficult to study the ICM at large
radii with X-rays.
Although several X-ray results have extended beyond
$R_{500}$ in individual clusters or small sets of clusters
\citep{george09, bautz09, reiprich09, simionescu11, walker12},
with current X-ray instrumentation it is infeasible
to extend such studies to large samples of clusters.
Consequently, two groups have used a hybrid approach
with X-ray data at small radii ($R \lesssim R_{500}$) and simulations 
at large radii ($R \gtrsim R_{500}$) in order to constrain the average
pressure profile at all relevant radial scales \citep[A10]{nagai07}.
These X-ray and simulation-based results imply that cluster pressure
profiles are approximately universal over a wide range of masses
and radial scales,
with low intrinsic scatter that is minimized near 0.5$R_{500}$
at $\lesssim 20$\% \citep[e.g., A10, ][]
{nagai07, sun11}.

The Sunyaev-Zel'dovich (SZ) effect signal \citep{sunyaev72}, which is proportional
to the density of the ICM and therefore falls more
slowly with radius compared to the X-ray brightness,
can be exploited to study the ICM pressure at large radii. 
Although some initial studies using \emph{WMAP}
SZ data showed large inconsistencies
with the established X-ray results \citep{lieu06, bielby07},
recent results have shown that SZ data from
\emph{WMAP}, \emph{Planck}, and ground-based receivers
provide a picture of the ICM that is consistent with X-ray
measurements at current observational precision 
\citep{plagge10, melin11, planck11_x, komatsu11, bonamente11}.
In particular, the South Pole Telescope (SPT) was able to measure
the average SZ pressure profile out to $\simeq 2R_{500}$
for a sample of 15 clusters \citep{plagge10},
finding results that were similar to previous X-ray/simulation
results at small/large radii (e.g., the sample of 31 REXCESS clusters
studied by A10).
Recently, a combination of \emph{XMM-Newton}
X-ray data at small radii
and \emph{Planck} SZ effect data
at large radii was used to constrain 
the average pressure profile out to $3R_{500}$
for a sample of 62 
\emph{Planck}-selected clusters (P12).
These results were again largely consistent with previous
analyses, and the X-ray and SZ data agreed quite well
in the overlapping region at intermediate radii.
Altogether, X-ray and SZ data, along with simulations, are 
converging to a uniform picture of the average cluster pressure 
profile over a wide range of angular scales.

This manuscript is arranged as follows.
In Section~\ref{sec:sample} we describe our sample of
45 massive clusters, and in Section~\ref{sec:reduction}
we provide the details of our SZ and X-ray data
reduction.
We then present our method for deprojecting pressure
profiles from our spatially-filtered SZ images
in Section~\ref{sec:deproj}.
In Section~\ref{sec:gNFW}, we describe parametric
fits to these deprojected profiles, and compare the
results of our fits to the results from
a range of previous analyses.
Then, in Section~\ref{sec:mass}, we use a Gaussian
process formalism to simultaneously constrain
the pressure-profile mass scaling, the ensemble
mean profile, and the radius-dependent intrinsic
scatter about this mean profile.
Finally, we provide a summary of our results
in Section~\ref{sec:summary}.

\section{Cluster Sample}
\label{sec:sample}

Between 2006 November and 2012 March we used Bolocam
to image the SZ signals from a sample of 45 clusters
that have {\it Chandra} X-ray exposures (hereafter the Bolocam X-ray/SZ
or BOXSZ sample). The Bolocam SZ observations and some general properties 
of the BOXSZ sample are
summarized in Table~\ref{tab:cluster_sample},
and the details of the {\it Chandra} X-ray data and analysis are given
in Tables~\ref{tab:chandra} and \ref{tab:cluster_sample_2}.
The BOXSZ sample contains two previously defined subsamples, the 25 object
Cluster Lensing and Supernova Survey with Hubble (CLASH)
sample \citep{postman12} and the 12 cluster MACS high-$z$ sample \citep{ebeling07}.
The remaining clusters were selected in an \emph{ad hoc} manner, with a general emphasis
on massive and/or high redshift systems. The clusters span the redshift range
$0.15 \le z \le 0.89$, with a median redshift of 0.42 and more than 60\% of
the sample lying between $0.35 \le z \le 0.59$.
The clusters are among the most massive known, with a median
X-ray derived mass of $M_{500} = 9 \times 10^{14}$~M$_{\odot}$
assuming a constant gas mass fraction
(we computed these masses using a reference 
$h = 0.7$, $\Omega_m = 0.3$ flat $\Lambda$CDM cosmology,
and this same cosmology was used to calculate all other
physical quantities presented in this manuscript).
The clusters in the BOXSZ sample span a range of dynamical states, from relaxed systems
with well defined cool-cores like Abell 1835 \citep{peterson01, schmidt01} to clusters
undergoing major merger events like MACS J0717.5 \citep{ebeling01, edge03, mroczkowski12}.

\begin{deluxetable*}{ccccccc} 
  \tablewidth{0pt}
   \tablecaption{Bolocam X-ray/SZ (BOXSZ) Cluster Sample}
   \tablehead{\colhead{cluster} & \colhead{redshift} &
     \colhead{RA} & \colhead{dec} &
     \colhead{obs. time} & \colhead{noise} &
     \colhead{peak S/N} \\
     \colhead{} & \colhead{} &
     \colhead{J2000} & \colhead{J2000} &
     \colhead{hours} & \colhead{\uKamin} &
     \colhead{}}
   \startdata
Abell 2204 & 0.151$^a$ & 16:32:47.2 & $+$05:34:33 &  12.7 &  18.5 &  22.3\\
Abell 383 & 0.188$^b$ & 02:48:03.3 & $-$03:31:46 &  24.3 &  18.9 & \phn  9.6\\
Abell 209 & 0.206$^a$ & 01:31:53.1 & $-$13:36:48 &  17.8 &  22.3 &  13.9\\
Abell 963 & 0.206$^a$ & 10:17:03.6 & $+$39:02:52 &  11.0 &  35.7 & \phn  8.3\\
Abell 1423 & 0.213$^a$ & 11:57:17.4 & $+$33:36:40 &  11.5 &  31.7 & \phn  5.8\\
Abell 2261 & 0.224$^a$ & 17:22:27.0 & $+$32:07:58 &  17.5 &  15.9 &  10.2\\
Abell 2219 & 0.228$^a$ & 16:40:20.3 & $+$46:42:30 & \phn  6.3 &  39.6 &  11.1\\
Abell 267 & 0.230$^a$ & 01:52:42.2 & $+$01:00:30 &  20.7 &  23.0 & \phn  9.6\\
RX J2129.6 & 0.235$^a$ & 21:29:39.7 & $+$00:05:18 &  16.0 &  23.7 & \phn  8.0\\
Abell 1835 & 0.253$^a$ & 14:01:01.9 & $+$02:52:40 &  14.0 &  16.2 &  15.7\\
Abell 697 & 0.282$^a$ & 08:42:57.6 & $+$36:21:57 &  14.3 &  17.4 &  22.6\\
Abell 611 & 0.288$^b$ & 08:00:56.8 & $+$36:03:26 &  18.7 &  25.0 &  10.8\\
MS 2137 & 0.313$^b$ & 21:40:15.1 & $-$23:39:40 &  12.8 &  27.3 & \phn  6.5\\
Abell S1063 & 0.348$^c$ & 22:48:44.8 & $-$44:31:45 & \phn  5.5 &  48.6 &  10.2\\
MACS J1931.8 & 0.352$^a$ & 19:31:49.6 & $-$26:34:34 & \phn  7.5 &  28.7 &  10.1\\
MACS J1115.8 & 0.355$^a$ & 11:15:51.9 & $+$01:29:55 &  15.7 &  22.8 &  10.9\\
MACS J1532.9 & 0.363$^a$ & 15:32:53.8 & $+$30:20:59 &  14.8 &  22.3 & \phn  8.0\\
Abell 370 & 0.375$^d$ & 02:39:53.2 & $-$01:34:38 &  11.8 &  28.9 &  12.8\\
MACS J1720.3 & 0.387$^a$ & 17:20:16.7 & $+$35:36:23 &  16.8 &  23.5 &  10.6\\
ZWCL 0024 & 0.395$^e$ & 00:26:35.8 & $+$17:09:41 & \phn  8.3 &  26.6 & \phn  3.3\\
MACS J2211.7 & 0.396$^a$ & 22:11:45.9 & $-$03:49:42 & \phn  6.5 &  38.6 &  14.7\\
MACS J0429.6 & 0.399$^a$ & 04:29:36.0 & $-$02:53:06 &  17.0 &  24.1 & \phn  8.9\\
MACS J0416.1 & 0.420$^f$ & 04:16:08.8 & $-$24:04:14 & \phn  7.8 &  29.3 & \phn  8.5\\
MACS J0451.9 & 0.430$^c$ & 04:51:54.7 & $+$00:06:19 &  14.2 &  22.7 & \phn  8.1\\
MACS J1206.2 & 0.439$^a$ & 12:06:12.3 & $-$08:48:06 &  11.3 &  24.9 &  21.7\\
MACS J0417.5 & 0.443$^a$ & 04:17:34.3 & $-$11:54:27 & \phn  9.8 &  22.7 &  22.7\\
MACS J0329.6 & 0.450$^b$ & 03:29:41.5 & $-$02:11:46 &  10.3 &  22.5 &  12.1\\
MACS J1347.5 & 0.451$^a$ & 13:47:30.8 & $-$11:45:09 &  15.5 &  19.7 &  36.6\\
MACS J1311.0 & 0.494$^b$ & 13:11:01.7 & $-$03:10:40 &  14.2 &  22.5 & \phn  9.6\\
MACS J2214.9 & 0.503$^a$ & 22:14:57.3 & $-$14:00:11 & \phn  7.2 &  27.3 &  12.6\\
MACS J0257.1 & 0.505$^a$ & 02:57:09.1 & $-$23:26:04 & \phn  5.0 &  39.0 &  10.1\\
MACS J0911.2 & 0.505$^a$ & 09:11:10.9 & $+$17:46:31 & \phn  6.2 &  33.5 & \phn  4.8\\
MACS J0454.1 & 0.538$^a$ & 04:54:11.4 & $-$03:00:51 &  14.5 &  18.2 &  24.3\\
MACS J1423.8 & 0.543$^a$ & 14:23:47.9 & $+$24:04:43 &  21.7 &  22.3 & \phn  9.4\\
MACS J1149.5 & 0.544$^a$ & 11:49:35.4 & $+$22:24:04 &  17.7 &  24.0 &  17.4\\
MACS J0018.5 & 0.546$^a$ & 00:18:33.4 & $+$16:26:13 & \phn  9.8 &  21.0 &  15.7\\
MACS J0717.5 & 0.546$^a$ & 07:17:32.1 & $+$37:45:21 &  12.5 &  29.4 &  21.3\\
MS 2053 & 0.583$^c$ & 20:56:21.0 & $-$04:37:49 &  18.7 &  18.0 & \phn  5.1\\
MACS J0025.4 & 0.584$^a$ & 00:25:29.9 & $-$12:22:45 &  14.3 &  19.7 &  12.3\\
MACS J2129.4 & 0.589$^a$ & 21:29:25.7 & $-$07:41:31 &  13.2 &  21.3 &  15.2\\
MACS J0647.7 & 0.591$^a$ & 06:47:49.7 & $+$70:14:56 &  11.7 &  22.0 &  14.4\\
MACS J0744.8 & 0.698$^a$ & 07:44:52.3 & $+$39:27:27 &  16.3 &  20.6 &  13.3\\
MS 1054 & 0.831$^g$ & 10:56:58.5 & $-$03:37:34 &  18.3 &  13.9 &  17.4\\
CL J0152.7 & 0.833$^a$ & 01:52:41.1 & $-$13:58:07 & \phn  9.3 &  23.4 &  10.2\\
CL J1226.9 & 0.888$^a$ & 12:26:57.9 & $+$33:32:49 &  11.8 &  22.9 &  13.0
   \enddata
   \tablecomments{The Bolocam X-ray/SZ (BOXSZ) cluster sample of 45 objects. The columns give the name,
   redshift, X-ray centroid coordinates (J2000), total Bolocam integration time, 
   RMS noise level of the SZ images, and peak S/N
   in the optimally filtered images (see \citealt{sayers12b} for details
   of how the peak S/N and optimal filter are determined).
   The superscripts denote the reference for the redshifts, with
   $a$) \citet{mantz10_ii},  
   $b$) \citet{allen08},
   $c$) \citet{maughan12},
   $d$) \citet{richard10},
   $e$) \citet{jee07},
   $f$) \citet{christensen12}, and
   $g$) \citet{tran07}.}
   \label{tab:cluster_sample}
\end{deluxetable*}

\begin{deluxetable}{ccccccc}
  \tablewidth{\columnwidth}
   \tablecaption{Details of the {\it Chandra} ACIS observations used in 
     this work.}
   \tablehead{\colhead{cluster} & \multicolumn{3}{c}{date} &
     \colhead{ObsID} & \colhead{mode} &
     \colhead{exp. (ks)}}
   \startdata
Abell 383            &   2000  &  Sep  &  08  &  524    &  VFAINT  &  \phn8.2  \\
                     &   2000  &  Nov  &  16  &  2320   &  VFAINT  &  16.8  \\
                     &   2000  &  Nov  &  16  &  2321   &  FAINT   &  17.2  \\
Abell 611            &   2001  &  Nov  &  03  &  3194   &  VFAINT  &  33.8  \\
Abell S1063                &   2004  &  May  &  17  &  4966   &  VFAINT  &  23.8  \\
Abell 370            &   1999  &  Oct  &  22  &  515    &  FAINT   &  67.8  \\
                     &   2006  &  Nov  &  26  &  7715   &  VFAINT  &  \phn6.6  \\
ZWCL 0024            &   2000  &  Sep  &  06  &  929    &  FAINT   &  33.8  \\
                     &   2006  &  Nov  &  28  &  7717   &  VFAINT  &  \phn7.0  \\
MACS J0416.1  &   2009  &  Jun  &  07  &  10446  &  VFAINT  &  14.3  \\
MACS J0451.9    &   2005  &  Jan  &  08  &  5815   &  VFAINT  &  \phn9.6  \\
MACS J0329.6  &   2001  &  Nov  &  25  &  3257   &  VFAINT  &  \phn8.1  \\
                     &   2002  &  Dec  &  24  &  3582   &  VFAINT  &  19.8  \\
                     &   2004  &  Dec  &  06  &  6108   &  VFAINT  &  35.3  \\
                     &   2006  &  Dec  &  03  &  7719   &  VFAINT  &  \phn7.1  \\
MACS J1311.0  &   2002  &  Dec  &  15  &  3258   &  VFAINT  &  14.9  \\
                     &   2005  &  Apr  &  20  &  6110   &  VFAINT  &  63.0  \\
                     &   2007  &  Mar  &  03  &  7721   &  VFAINT  &  \phn6.6  \\
                     &   2007  &  Dec  &  09  &  9381   &  VFAINT  &  24.1  \\
MS 2053        &   2000  &  May  &  13  &  551    &  FAINT   &  37.7  \\
                     &   2001  &  Oct  &  07  &  1667   &  VFAINT  &  36.0  \\
MS 1054        &   2000  &  Apr  &  21  &  512    &  FAINT   &  75.8  \\
CL J0152.7     &   2000  &  Sep  &  08  &  913    &  FAINT   &  29.2  \\
CL J1226.9       &   2000  &  Jul  &  31  &  932    &  VFAINT  &  \phn9.8  \\
                     &   2003  &  Jan  &  27  &  3180   &  VFAINT  &  26.1  \\
                     &   2004  &  Aug  &  07  &  5014   &  VFAINT  &  25.3 
\enddata
\tablecomments{Cluster name, {\it Chandra} observation date, observation ID number, 
  observing mode used, and clean exposure time for clusters presented in this
  work which did not appear in \citet{mantz10_ii}.
  We refer the reader to that work for details of the other {\it Chandra} observations.}
\label{tab:chandra}
\end{deluxetable}

Based on previous results that show the projected X-ray luminosity ratio 
is an accurate indicator of cool-core clusters 
\citep{mantz09, bohringer10},
we define a cool-core subsample of the BOXSZ sample as those clusters with
a projected X-ray luminosity ratio
\begin{equation} \label{eq:Lrat}
L_\mathrm{rat} = \frac{L(R < 0.05R_{500})}
{L(R < R_{500})} \ge 0.17.
\end{equation}
According to this definition, 17/45 of the BOXSZ clusters are cool-core systems
(see Table~\ref{tab:cluster_sample_2}).
We note that these cool-core systems are in general at the low redshift end
of our sample, with a median redshift of $z=0.36$.
We speculate that this result is due at least in part 
to the fact that the cool-core fraction
of clusters drops with increasing redshift (e.g., \citealt{vikhlinin06b, santos10}), 
although selection effects may also play 
a role due to the \emph{ad hoc} manner in which the BOXSZ sample was chosen.

\begin{deluxetable*}{cccccccccc} 
  \tablewidth{0pt}
   \tablecaption{X-ray Properties of the BOXSZ Sample}
   \tablehead{\colhead{cluster} & \colhead{$R_{500}$} &
     \colhead{$L_{500}$} & \colhead{$M_{500}$} &
     \colhead{$kT$} & \colhead{$P_{500}$} &
     \colhead{$L_\mathrm{rat}$} & \colhead{$w_{500}$} &
     \colhead{cool-core} & \colhead{disturbed} \\
     \colhead{} & \colhead{Mpc} &
     \colhead{$10^{44}$ erg/s} & \colhead{$10^{14}$ M$_{\odot}$} &
     \colhead{keV} & \colhead{$10^{-3}$ keV/cm$^{3}$} &
     \colhead{} & \colhead{$10^{-2}$} &
     \colhead{} & \colhead{}}
   \startdata
Abell 2204 & $1.46 \pm 0.07$ & $ 17.9 \pm   1.6$ & $ 10.3 \pm   1.5$ & $\phn  8.6 \pm   0.6$ & $\phn 4.56 \pm  1.26$ & $0.35 \pm 0.07$ & $  0.13 \pm   0.04$ &  $\checkmark$  &   \\ 
Abell 383 & $1.11 \pm 0.06$ & $\phn  6.0 \pm   0.2$ & $\phn  4.7 \pm   0.8$ & $\phn  5.4 \pm   0.2$ & $\phn 2.85 \pm  0.87$ & $0.28 \pm 0.03$ & $  0.19 \pm   0.03$ &  $\checkmark$  &   \\ 
Abell 209 & $1.53 \pm 0.08$ & $\phn  8.6 \pm   0.3$ & $ 12.6 \pm   1.9$ & $\phn  8.2 \pm   0.7$ & $\phn 5.64 \pm  1.60$ & $0.07 \pm 0.02$ & $  0.50 \pm   0.17$ &   &   \\ 
Abell 963 & $1.25 \pm 0.06$ & $\phn  6.5 \pm   0.2$ & $\phn  6.8 \pm   1.0$ & $\phn  6.1 \pm   0.3$ & $\phn 3.74 \pm  1.04$ & $0.15 \pm 0.02$ & $  0.22 \pm   0.11$ &   &   \\ 
Abell 1423 & $1.35 \pm 0.10$ & $\phn  6.2 \pm   0.4$ & $\phn  8.7 \pm   2.0$ & $\phn  5.8 \pm   0.6$ & $\phn 4.45 \pm  1.67$ & $0.13 \pm 0.03$ & $  0.76 \pm   0.19$ &   &   \\ 
Abell 2261 & $1.59 \pm 0.09$ & $ 12.0 \pm   0.4$ & $ 14.4 \pm   2.6$ & $\phn  6.1 \pm   0.3$ & $\phn 6.32 \pm  2.02$ & $0.20 \pm 0.02$ & $  0.85 \pm   0.08$ &  $\checkmark$  &   \\ 
Abell 2219 & $1.74 \pm 0.08$ & $ 15.5 \pm   0.8$ & $ 18.9 \pm   2.5$ & $ 10.9 \pm   0.5$ & $\phn 7.62 \pm  1.98$ & $0.07 \pm 0.02$ & $  0.18 \pm   0.13$ &   &   \\ 
Abell 267 & $1.22 \pm 0.07$ & $\phn  5.8 \pm   0.2$ & $\phn  6.6 \pm   1.1$ & $\phn  7.1 \pm   0.7$ & $\phn 3.79 \pm  1.15$ & $0.08 \pm 0.02$ & $  2.68 \pm   1.26$ &   &  $\checkmark$  \\ 
RX J2129.6 & $1.28 \pm 0.07$ & $\phn  9.9 \pm   0.5$ & $\phn  7.7 \pm   1.2$ & $\phn  6.3 \pm   0.6$ & $\phn 4.23 \pm  1.23$ & $0.25 \pm 0.03$ & $  0.52 \pm   0.14$ &  $\checkmark$  &   \\ 
Abell 1835 & $1.49 \pm 0.06$ & $ 21.1 \pm   0.6$ & $ 12.3 \pm   1.4$ & $\phn  9.0 \pm   0.2$ & $\phn 5.94 \pm  1.39$ & $0.36 \pm 0.02$ & $  0.23 \pm   0.02$ &  $\checkmark$  &   \\ 
Abell 697 & $1.65 \pm 0.09$ & $ 14.4 \pm   0.8$ & $ 17.1 \pm   2.9$ & $ 10.9 \pm   1.1$ & $\phn 7.72 \pm  2.36$ & $0.08 \pm 0.02$ & $  0.60 \pm   0.45$ &   &   \\ 
Abell 611 & $1.24 \pm 0.06$ & $\phn  7.5 \pm   0.4$ & $\phn  7.4 \pm   1.1$ & $\phn  6.8 \pm   0.3$ & $\phn 4.45 \pm  1.25$ & $0.16 \pm 0.03$ & $  0.56 \pm   0.10$ &   &   \\ 
MS 2137 & $1.06 \pm 0.04$ & $ 11.1 \pm   0.4$ & $\phn  4.7 \pm   0.6$ & $\phn  4.7 \pm   0.4$ & $\phn 3.42 \pm  0.87$ & $0.40 \pm 0.03$ & $  0.39 \pm   0.05$ &  $\checkmark$  &   \\ 
Abell S1063 & $1.76 \pm 0.09$ & $ 30.8 \pm   1.6$ & $ 22.2 \pm   3.4$ & $ 10.9 \pm   0.5$ & $10.14 \pm  2.90$ & $0.16 \pm 0.04$ & $  0.75 \pm   0.15$ &   &   \\ 
MACS J1931.8 & $1.34 \pm 0.07$ & $ 19.7 \pm   1.0$ & $\phn  9.9 \pm   1.6$ & $\phn  7.5 \pm   1.4$ & $\phn 5.95 \pm  1.77$ & $0.40 \pm 0.04$ & $  0.35 \pm   0.09$ &  $\checkmark$  &   \\ 
MACS J1115.8 & $1.28 \pm 0.06$ & $ 14.5 \pm   0.5$ & $\phn  8.6 \pm   1.2$ & $\phn  9.2 \pm   1.0$ & $\phn 5.45 \pm  1.47$ & $0.28 \pm 0.02$ & $  0.27 \pm   0.05$ &  $\checkmark$  &   \\ 
MACS J1532.9 & $1.31 \pm 0.08$ & $ 19.8 \pm   0.7$ & $\phn  9.5 \pm   1.7$ & $\phn  6.8 \pm   1.0$ & $\phn 5.89 \pm  1.87$ & $0.38 \pm 0.03$ & $  0.28 \pm   0.15$ &  $\checkmark$  &   \\ 
Abell 370 & $1.40 \pm 0.08$ & $\phn  8.6 \pm   0.4$ & $ 11.7 \pm   2.1$ & $\phn  7.3 \pm   0.5$ & $\phn 6.89 \pm  2.19$ & $0.04 \pm 0.01$ & $  4.90 \pm   2.00$ &   &  $\checkmark$  \\ 
MACS J1720.3 & $1.14 \pm 0.07$ & $ 10.2 \pm   0.4$ & $\phn  6.3 \pm   1.1$ & $\phn  7.9 \pm   0.7$ & $\phn 4.65 \pm  1.45$ & $0.26 \pm 0.02$ & $  0.24 \pm   0.06$ &  $\checkmark$  &   \\ 
ZWCL 0024 & $1.00 \pm 0.11$ & $\phn  2.3 \pm   0.1$ & $\phn  4.4 \pm   1.6$ & $\phn  5.9 \pm   0.9$ & $\phn 3.70 \pm  1.89$ & $0.10 \pm 0.03$ & $  2.53 \pm   0.41$ &   &  $\checkmark$  \\ 
MACS J2211.7 & $1.61 \pm 0.07$ & $ 24.0 \pm   1.2$ & $ 18.1 \pm   2.5$ & $ 14.0 \pm   2.7$ & $\phn 9.52 \pm  2.55$ & $0.19 \pm 0.03$ & $  0.88 \pm   0.13$ &  $\checkmark$  &   \\ 
MACS J0429.6 & $1.10 \pm 0.05$ & $ 10.9 \pm   0.6$ & $\phn  5.8 \pm   0.8$ & $\phn  8.3 \pm   1.6$ & $\phn 4.48 \pm  1.20$ & $0.33 \pm 0.04$ & $  0.39 \pm   0.07$ &  $\checkmark$  &   \\ 
MACS J0416.1 & $1.27 \pm 0.15$ & $\phn  8.1 \pm   0.5$ & $\phn  9.1 \pm   2.0$ & $\phn  8.2 \pm   1.0$ & $\phn 6.25 \pm  2.28$ & $0.04 \pm 0.02$ & $  2.02 \pm   1.06$ &   &  $\checkmark$  \\ 
MACS J0451.9 & $1.12 \pm 0.06$ & $\phn  6.7 \pm   0.5$ & $\phn  6.3 \pm   1.1$ & $\phn  6.7 \pm   1.0$ & $\phn 4.97 \pm  1.55$ & $0.08 \pm 0.03$ & $  1.93 \pm   0.80$ &   &  $\checkmark$  \\ 
MACS J1206.2 & $1.61 \pm 0.08$ & $ 21.1 \pm   1.1$ & $ 19.2 \pm   3.0$ & $ 10.7 \pm   1.3$ & $10.59 \pm  3.07$ & $0.15 \pm 0.03$ & $  0.72 \pm   0.11$ &   &   \\ 
MACS J0417.5 & $1.69 \pm 0.07$ & $ 29.1 \pm   1.5$ & $ 22.1 \pm   2.7$ & $\phn  9.5 \pm   1.1$ & $11.70 \pm  2.88$ & $0.19 \pm 0.03$ & $  3.01 \pm   0.07$ &  $\checkmark$  &  $\checkmark$  \\ 
MACS J0329.6 & $1.19 \pm 0.06$ & $ 13.4 \pm   0.4$ & $\phn  7.9 \pm   1.3$ & $\phn  6.3 \pm   0.3$ & $\phn 5.96 \pm  1.79$ & $0.33 \pm 0.02$ & $  1.40 \pm   0.26$ &  $\checkmark$  &  $\checkmark$  \\ 
MACS J1347.5 & $1.67 \pm 0.08$ & $ 42.2 \pm   1.1$ & $ 21.7 \pm   3.0$ & $ 10.8 \pm   0.8$ & $11.71 \pm  3.13$ & $0.39 \pm 0.02$ & $  0.59 \pm   0.04$ &  $\checkmark$  &   \\ 
MACS J1311.0 & $0.93 \pm 0.04$ & $\phn  7.5 \pm   0.2$ & $\phn  3.9 \pm   0.5$ & $\phn  6.0 \pm   0.3$ & $\phn 3.99 \pm  1.01$ & $0.19 \pm 0.01$ & $  0.22 \pm   0.08$ &  $\checkmark$  &   \\ 
MACS J2214.9 & $1.39 \pm 0.08$ & $ 13.9 \pm   0.6$ & $ 13.2 \pm   2.3$ & $\phn  9.6 \pm   0.8$ & $\phn 9.12 \pm  2.85$ & $0.10 \pm 0.02$ & $  1.30 \pm   0.29$ &   &  $\checkmark$  \\ 
MACS J0257.1 & $1.20 \pm 0.06$ & $ 12.1 \pm   0.5$ & $\phn  8.5 \pm   1.3$ & $\phn  9.9 \pm   0.9$ & $\phn 6.82 \pm  1.95$ & $0.12 \pm 0.02$ & $  0.46 \pm   0.13$ &   &   \\ 
MACS J0911.2 & $1.22 \pm 0.06$ & $\phn  7.5 \pm   0.3$ & $\phn  9.0 \pm   1.2$ & $\phn  6.6 \pm   0.6$ & $\phn 7.09 \pm  1.85$ & $0.05 \pm 0.01$ & $  0.89 \pm   0.64$ &   &   \\ 
MACS J0454.1 & $1.31 \pm 0.06$ & $ 15.7 \pm   0.6$ & $ 11.5 \pm   1.5$ & $\phn  9.1 \pm   0.5$ & $\phn 8.79 \pm  2.26$ & $0.07 \pm 0.01$ & $  2.27 \pm   1.50$ &   &  $\checkmark$  \\ 
MACS J1423.8 & $1.09 \pm 0.05$ & $ 14.0 \pm   0.5$ & $\phn  6.6 \pm   0.9$ & $\phn  6.9 \pm   0.3$ & $\phn 6.12 \pm  1.62$ & $0.37 \pm 0.03$ & $  0.31 \pm   0.15$ &  $\checkmark$  &   \\ 
MACS J1149.5 & $1.53 \pm 0.08$ & $ 17.2 \pm   0.7$ & $ 18.7 \pm   3.0$ & $\phn  8.5 \pm   0.6$ & $12.28 \pm  3.62$ & $0.05 \pm 0.01$ & $  1.64 \pm   1.23$ &   &  $\checkmark$  \\ 
MACS J0018.5 & $1.47 \pm 0.08$ & $ 18.0 \pm   0.9$ & $ 16.5 \pm   2.5$ & $\phn  9.1 \pm   0.4$ & $11.33 \pm  3.22$ & $0.06 \pm 0.02$ & $  0.67 \pm   0.14$ &   &   \\ 
MACS J0717.5 & $1.69 \pm 0.06$ & $ 25.0 \pm   0.9$ & $ 24.9 \pm   2.7$ & $ 11.8 \pm   0.5$ & $14.90 \pm  3.39$ & $0.05 \pm 0.01$ & $  2.55 \pm   1.26$ &   &  $\checkmark$  \\ 
MS 2053 & $0.82 \pm 0.06$ & $\phn  2.8 \pm   0.1$ & $\phn  3.0 \pm   0.5$ & $\phn  4.4 \pm   0.6$ & $\phn 3.86 \pm  1.17$ & $0.07 \pm 0.02$ & $  1.02 \pm   0.31$ &   &  $\checkmark$  \\ 
MACS J0025.4 & $1.12 \pm 0.04$ & $\phn  9.1 \pm   0.4$ & $\phn  7.6 \pm   0.9$ & $\phn  6.5 \pm   0.5$ & $\phn 7.18 \pm  1.73$ & $0.03 \pm 0.01$ & $  0.65 \pm   0.50$ &   &   \\ 
MACS J2129.4 & $1.25 \pm 0.06$ & $ 13.7 \pm   0.6$ & $ 10.6 \pm   1.4$ & $\phn  8.6 \pm   0.7$ & $\phn 9.03 \pm  2.34$ & $0.08 \pm 0.02$ & $  1.51 \pm   0.69$ &   &  $\checkmark$  \\ 
MACS J0647.7 & $1.26 \pm 0.06$ & $ 14.1 \pm   0.6$ & $ 10.9 \pm   1.6$ & $ 11.5 \pm   1.1$ & $\phn 9.23 \pm  2.57$ & $0.10 \pm 0.02$ & $  0.62 \pm   0.29$ &   &   \\ 
MACS J0744.8 & $1.26 \pm 0.06$ & $ 18.9 \pm   0.6$ & $ 12.5 \pm   1.6$ & $\phn  8.1 \pm   0.4$ & $11.99 \pm  3.05$ & $0.16 \pm 0.02$ & $  1.60 \pm   0.11$ &   &  $\checkmark$  \\ 
MS 1054 & $1.07 \pm 0.07$ & $ 12.4 \pm   0.7$ & $\phn  9.0 \pm   1.3$ & $ 12.0 \pm   1.4$ & $11.90 \pm  3.28$ & $0.02 \pm 0.01$ & $  6.62 \pm   2.47$ &   &  $\checkmark$  \\ 
CL J0152.7 & $0.97 \pm 0.26$ & $\phn  7.3 \pm   0.6$ & $\phn  7.8 \pm   3.0$ & $\phn  6.5 \pm   0.9$ & $10.86 \pm  5.74$ & $0.01 \pm 0.01$ & $  8.22 \pm   1.02$ &   &  $\checkmark$  \\ 
CL J1226.9 & $1.00 \pm 0.05$ & $ 14.0 \pm   0.5$ & $\phn  7.8 \pm   1.1$ & $ 12.0 \pm   1.3$ & $11.84 \pm  3.21$ & $0.10 \pm 0.02$ & $  0.95 \pm   0.31$ &   &   
   \enddata
   \tablecomments{Relevant X-ray derived properties of the BOXSZ cluster sample.
   The first four columns provide $R_{500}$, $L_{500}$, $M_{500}$, and $kT$, computed as described in 
   \citet{mantz10_ii}. The fifth column gives $P_{500}$, computed from $M_{500}$ and $z$
   according to Equation~\ref{eqn:p500}, given in Section~\ref{sec:deproj}.
   The sixth column gives the ratio of the projected X-ray luminosity within 0.05$R_{500}$
   to the X-ray luminosity within $R_{500}$ \citep{mantz09}. The seventh column gives the 
   centroid shift parameter within $R_{500}$, computed according to the 
   procedure described in \citet{maughan08, maughan12}.
   We denote cool-core clusters as those having an $L_\mathrm{rat} \ge 0.17$, and 
   we denote disturbed clusters as those having $w_{500} \ge 0.01$.}
   \label{tab:cluster_sample_2}
\end{deluxetable*}

We define a disturbed subsample of the BOXSZ sample as those clusters with 
an X-ray centroid
shift parameter of $w_{500} \ge 0.01$
(see Section~\ref{sec:xray} for a full description of how we compute $w_{500}$). 
The centroid shift parameter 
is widely used to classify disturbed systems 
(e.g., \citealt{maughan08, pratt09, maughan12}),
and we adopt the same threshold ($w_{500} \ge 0.01$) as \citet{pratt09}.
Based on this criteria, 16/45 clusters in the BOXSZ sample are disturbed, and the
disturbed systems are generally at the high redshift end of the
full sample with a median redshift of $z = 0.52$.
This result is not surprising given that the cool-core systems are generally
at low redshift and cool cores are a good indicator that the
cluster is not disturbed (only 2/45 clusters in the BOXSZ sample have both
a cool core and are disturbed).

\begin{figure}
  \includegraphics[width=\columnwidth]{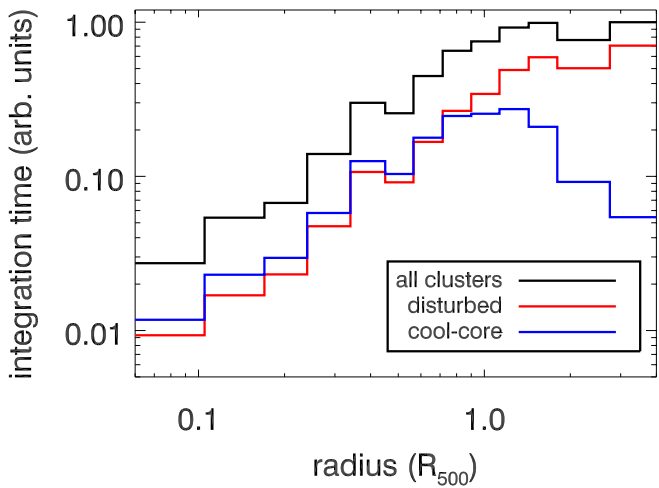}
  \caption{Total Bolocam integration time for BOXSZ clusters
  as a function of scaled radius for the same 13 radial bins
  given in the left column of Table~\ref{tab:deprojection}.
  The black line denotes
  the full BOXSZ sample, the red line denotes the disturbed subsample,
  and the blue line denotes the cool-core subsample.}
  \label{fig:cluster_radii}
\end{figure}

This redshift asymmetry between the cool-core and disturbed subsamples
within the BOXSZ sample has an impact on our analysis of the average pressure profile 
of the sample. As we describe in Section~\ref{sec:deproj}, we scale the radial
coordinate of each cluster by $R_{500}$,
since the shape
of the pressure profiles is expected to be self-similar after scaling by this radius.
However, the angular dynamic range of our data is limited
by Bolocam's 58~arcsec full-width at half-maximum (FWHM) point-spread
function (PSF) and by the 14~arcmin size of the Bolocam images
(i.e., our SZ data are sensitive to a minimum radius of $\simeq 29$~arcsec and
a maximum radius of $\simeq 10$~arcmin). Since the value of $R_{500}$
in physical units is fairly constant over our sample ($\simeq 1.5$~Mpc), this means
that the value of $R_{500}$ varies significantly in angular size
over the redshift range of our sample, from 2~arcmin to 9~arcmin. 
This can be clearly seen
in Figure~\ref{fig:cluster_radii}, where we show the Bolocam integration time
as a function of scaled radius for the BOXSZ sample.
On average, we obtained longer integrations for the preferentially
high redshift disturbed clusters,
and this compensates for their smaller angular size compared to the 
preferentially low redshift cool-core
clusters to produce nearly equal integration times at 
$R \lesssim R_{500}$.
However, the increased integration time and smaller angular
sizes of the disturbed clusters results in significantly more integration
time outside $R_{500}$ compared to the cool-core clusters.
Consequently, the average SZ pressure profile of the BOXSZ sample is constrained
by an above average amount of data from disturbed systems
at larger radii.
At these large radii, data and simulations indicate that there is little
or no difference in the pressure profiles based on morphological 
classification, so we expect that any resulting bias in our results
will be minor \citep[A10]{borgani04, nagai07, piffaretti08}. 
We address this issue in more detail in 
Section~\ref{sec:gNFW}, where we examine the average pressure profiles
of our cool-core and disturbed subsamples separately and find that
they are consistent outside of $\simeq 0.15R_{500}$ given our
measurement uncertainties.

\section{Data Reduction}
\label{sec:reduction}

  \subsection{Bolocam}
    \label{sec:Bolocam_reduction}

    Bolocam is a 144-element bolometric imaging photometer, and 
    from 2003-2012 it served as the long-wavelength facility camera for
    the Caltech Submillimeter Observatory (CSO).
    Bolocam covers an eight-arcminute-diameter circular field of view (FOV)
    and has PSFs with 58~arcsec FWHMs
    \citep{glenn98, haig04}.
    The SZ-emission-weighted band center of our data is 140~GHz.
    All of the cluster images were obtained by scanning the CSO
    in a Lissajous pattern \citep{kovacs06}, with an amplitude
    of 4~arcmin and an average scan speed of $\simeq 4$~arcmin/sec.
    These scans result in images with tapered coverage extending to
    a radius of $\simeq 12$~arcmin, with the coverage dropping to
    half its peak value at a radius of $\simeq 5$~arcmin.
    For ease of analysis, we have made $14 \times 14$~arcmin
    square maps for each cluster.

    Our Bolocam data reduction largely followed the procedure described
    in detail in \citet{sayers11}, and we therefore briefly summarize
    that procedure below.
    First, we use frequent observations of bright compact objects to
    obtain pointing corrections accurate to 5~arcsec.
    Additionally, we made nightly observations of Uranus, Neptune,
    and/or other secondary calibrators to obtain flux calibration
    accurate to 5\% \citep{griffin93, sandell94, sayers12}.
    To remove noise from atmospheric fluctuations, we 
    subtract the FOV-average signal at each time sample in the
    time-ordered data (TOD) and also high-pass filter the TOD
    at a characteristic frequency of 250~mHz.
    This process also attenuates the astronomical signals in our data,
    and we characterize this filtering as follows.
    First, for each observation, we insert a model cluster profile in our TOD, process
    these model-plus-data TOD through our reduction pipeline, and create a map.
    We then subtract the data-only map to produce a noiseless image
    of the model after going through our data processing pipeline.
    The result is compared to the original input model to obtain
    a complex-valued two-dimensional map-space transfer function.

    To characterize the non-astronomical noise in our images
    we form jackknife realizations of the data by multiplying
    a randomly selected subset of half of the data by $-1$
    prior to binning the data into a map.
    For each cluster we formed 1000 such jackknife maps.
    To each of these maps we then added a Gaussian random realization
    of the 140~GHz astronomical sky based on the power spectrum measurements
    made by the SPT, which cover
    all of the angular scales probed by our data
    \citep{keisler11, reichardt12}.
    Each of these 1000 astronomical signal realizations was processed
    through our data reduction pipeline so that it was filtered
    identically to our real data.
    We have verified that the above noise model is statistically
    equivalent to measurements of blank sky made with Bolocam \citep{sayers11}.

    In addition, for 11 of our clusters we have subtracted
    individual bright point sources selected from
    the NVSS 1.4~GHz catalog \citep{condon98}.
    We refer the reader to \citet{sayers_radio} for a full description of these sources,
    most of which are near the cluster centers.
    Since all of these sources are below our 140~GHz detection limit, 
    we have extrapolated spectral fits to 1.4 and 30~GHz data.
    In almost all cases, the uncertainty on these extrapolated flux
    densities is $\simeq 30$\%, limited by the intrinsic scatter
    in the extrapolation.
    We subtracted all of the sources with extrapolated flux densities $>0.5$~mJy.
    This source brightness threshold was chosen to ensure that contamination of the cluster signal from
    point sources is $<1$\%.
    To account for our uncertainty in the flux density of these
    subtracted point sources, we add a model of each point source,
    multiplied by a random value drawn from a Gaussian distribution
    described by our uncertainty on the extrapolated source flux density,
    to each of the 1000 noise realizations for a given cluster observation.

    Furthermore, we detect a total of 6 bright point sources in
    our 140~GHz data, all of which were subtracted for this analysis \citep{sayers_radio}.
    For these sources, we refit the point source model to each of
    our 1000 jackknife realizations and added a point source template
    to each of our 1000 noise realizations based on the dispersion of these fits.

  \subsection{X-ray data}
    \label{sec:xray}

    X-ray luminosities, temperatures and masses for the BOXSZ clusters appearing in \citet{mantz10_ii} 
    (hereafter M10) are 
    taken from that work. For the other clusters, 
    these quantities were derived from archival {\it Chandra} data following the 
    same procedure as was used in M10, and we refer the reader there for 
    full details. Briefly, the archival data were reprocessed 
    using {\tt ciao}\footnote{\tt http://cxc.harvard.edu/ciao/} (version 
    4.1.1; CALDB 4.1.2), including removal of bad pixels, corrections for 
    cosmic ray afterglows and charge transfer inefficiency, and application 
    of standard grade and status filters, using appropriate time-dependent 
    gain and calibration products. Soft-band surface brightness profiles 
    were extracted and were scaled by a global factor to agree with the 
    final {\it ROSAT} flux calibration\footnote{
      This scaling of the {\it Chandra} data
      is described in detail in Section 2.2.4 of M10
      and was motivated by the primary goal of that analysis, which was to 
      relate {\it Chandra}-derived masses to {\it ROSAT} survey fluxes.
      For consistency with the results of M10, we have
      retained this scaling in our present analysis.}.
    These profiles were then used to derive 
    cluster luminosity, projected luminosity ratio (Equation~\ref{eq:Lrat}), and gas mass. 
    Values of total mass and $R_{500}$ were 
    derived for a reference $h=0.7$, $\Omega_m=0.3$, flat $\Lambda$CDM 
    cosmology using the derived gas mass profiles and the universal gas mass 
    fraction measured by \citet{allen08}. Spectra were extracted from an 
    annulus covering radii between 0.15$R_{500}$ and $R_{500}$ and fit in 
    {\tt xspec}\footnote{\tt 
      http://heasarc.gsfc.nasa.gov/docs/xanadu/xspec/} to provide a global 
    temperature measurement. For the lowest redshift cluster (Abell 2204), 
    the luminosity and gas mass analyses used ROSAT Position Sensitive Proportional Counter data, 
    since the {\it Chandra} field of view does not comfortably encompass 
    $R_{500}$, and the average temperature measured from ASCA data by 
    \citet{horner99} was adopted.

    In addition, we used the {\it Chandra}
    data to constrain the centroid shift parameter
    $w_{500}$. This analysis was described in 
    \citet{maughan12}, and we refer the reader to that
    manuscript for additional details.
    Centroid shifts measure the standard deviation of the projected separation 
    between the X-ray peak and the centroid as a function of projected radius 
    $R_p < R_{500}$. 
    Following the same approach as \citet{poo06}, $w_{500}$ was computed from a 
    series of circular apertures with initial and final radii $R_p=R_{500}$ and 
    $R_p = 0.05R_{500}$ respectively, decreasing in size by $0.05R_{500}$ in each iteration. 
    To this end, we used background-subtracted images, appropriately divided by the 
    exposure map to eliminate instrumental artifacts such as chip gaps and vignetting. 
    All point sources were excluded from the analysis; however extended sources 
    were left untouched as these may be associated with some of the cluster substructure.

\section{Pressure Deprojections}
\label{sec:deproj}

In order to determine physical pressures from our SZ data
(which are measured in units of a CMB fluctuation temperature) we 
use the equations described by \citet{sunyaev72}
with the relativistic corrections given by \citet{itoh98}.
For the relativistic corrections, we have assumed that the 
ICM of each cluster is isothermal, with a temperature equal to the
spectroscopic X-ray temperature given in Table~\ref{tab:cluster_sample_2}.
Although the true temperature profiles are expected to vary
by factors of $\simeq 2$ over the radii probed by our data,
we note that the relativistic corrections are $\lesssim 10$\%
for these clusters in our observing band.
Therefore, even if the ICM temperature varies by a factor of two over 
the radial range probed by our data, then
our isothermal approximation will produce less than
a 5\% bias on the pressures that we derive.

Due to the high-pass filtering applied to the Bolocam data to remove atmospheric
noise, it is not possible to directly deproject pressure profiles
from the standard Bolocam images.
We have demonstrated the ability to deconvolve the effects of this high pass filter
\citep{sayers11},
and we are able to produce unbiased images that could in principal be deprojected.
However, to prevent unphysical numerical artifacts 
from appearing in the deconvolved images,
we would need to reduce the size of our images
from 14 to 10 arcmin.
Since we are interested in information about the pressure profiles at large
radii, we do not want to accept this loss of information.

Consequently, we compute our deprojected pressure profiles as follows.
First, we select a set of discrete radial points (referred to as radial
bins) at which we would like to compute the spherically symmetric 
deprojected pressure.
Next, we create a smooth and continuous pressure profile by connecting
these radial bins with a power-law interpolation.
This smooth pressure profile is then projected into a two 
dimensional SZ model.
For the projection, we assume that the profile is equal to zero
outside of $5R_{500}$, and we use a power-law extrapolation
to estimate the pressure profile beyond
our largest radial bin and inside $5R_{500}$.
We then filter this SZ model with both the signal transfer function
of our data and the Bolocam PSF, and then compare the
filtered SZ model to our data map.

\begin{figure}
  \includegraphics[width=\columnwidth]{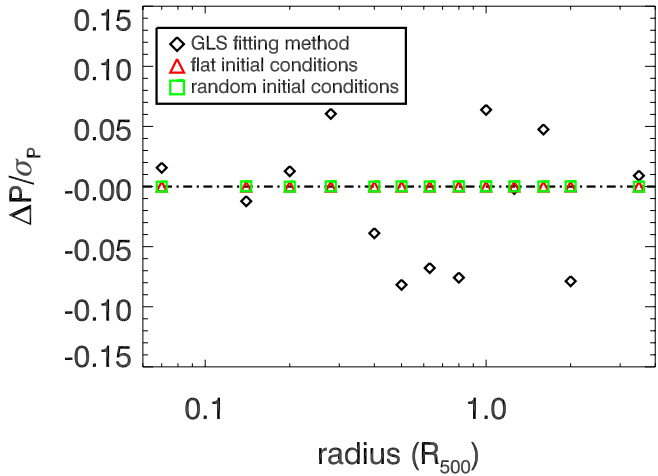}
  \caption{The GLS fitting bias divided by the uncertainty for each radial
    bin in the joint pressure deprojection of the full BOXSZ sample. 
    The black diamonds show the difference
    between the input pressure profile and the average recovered
    pressure profile using our GLS algorithm. The red triangles
    and green squares show the additional bias associated with
    choosing different starting values for the pressure bins relative to
    our default starting values equal to the best-fit model from A10
    (For the red triangles all of the starting values were set equal to
    $P_{500}$, and for the green squares all of the starting values
    were set equal to a randomly drawn value
    between 0 and $10P_{500}$). 
    In all cases, the bias is quasi-negligible when added in quadrature
    with the uncertainty (note that the typical S/N per bin is
    $\simeq 5$, so the absolute bias is $\simeq 2$\%).
    This result shows that our fitting method is approximately
    unbiased and that it is robust to our choice of initial conditions
    (i.e., it adequately explores the parameter space).}
  \label{fig:GLS}
\end{figure}

We determine the best fit pressure deprojections using 
the MPFITFUN generalized least squares (GLS)
fitting algorithm \citep{markwardt09} under the simplifying assumption that the noise covariance matrix
of our SZ maps is diagonal.
As described in \citet{sayers11}, a diagonal noise covariance matrix is a very
good, but not perfect, description of our SZ map data.
Consequently, we compute all of our uncertainties using the 1000 different noise
realizations described in Section~\ref{sec:Bolocam_reduction}.
First, we add a cluster model, equal to the best fit deprojection of our data,
separately to each of these 1000 noise realizations.
We then fit a pressure deprojection to each of these 1000 model-plus-noise
realizations, and quantify all of our uncertainties based on the spread
of these 1000 fits.
By design, this procedure fully accounts for all of the characteristics of
our noise to quantify our uncertainties for a 
deprojection of a cluster with a pressure
profile described by our best-fit deprojection.
One possible bias in this approach is that it only quantifies
the uncertainties for a particular cluster shape, and the true
cluster profile might differ from our best-fit profile.
Consequently, we tested this effect by computing uncertainties for 
our deprojection fits using both the best-fit
profile to our cluster sample (Section~\ref{sec:gNFW}) and the best-fit
profile found by A10 as the input shapes added to our 1000 noise realizations.
We find that the uncertainties recovered using an A10 profile differ
from the recovered uncertainties using our best-fit profile by
an rms of 7.7\%.
If the uncertainties were identical, then we would expect an rms
difference of 4.5\% due to our finite number of noise realizations.
This indicates that our uncertainties on the recovered deprojection
depend slightly on the exact shape of the cluster pressure profile.
However, since the true cluster pressure profile is likely to be
similar to our best-fit profile, and since the variation in
recovered uncertainties is similar to our precision in estimating
them due to our finite number of noise realizations, we do not
attempt to account for the cluster-shape dependence of our
recovered uncertainties in our analysis.
Finally, we have verified that our GLS algorithm adequately samples the parameter space
by showing that it is robust to our choice of initial conditions, 
and that we are able to recover an
input cluster candidate with minimal bias (see Figure~\ref{fig:GLS}).

Throughout this work we compare and average the pressure profiles from
multiple clusters.
Therefore, prior to any deprojection,
we have scaled the radial coordinate of each cluster by $R_{500}$
and the pressure amplitude by 
\begin{eqnarray}
  P_{500} = \left(3.68 \times 10^{-3} \frac{\mathrm{keV}}{\mathrm{cm}^3}\right) 
    \left(\frac{M_{500}}{10^{15} M_{\odot}} \right)^{\alpha_{P}} E(z)^{8/3},
  \label{eqn:p500}
\end{eqnarray}
where $E^2(z) = \Omega_M(1+z)^3 + \Omega_{\Lambda}$ and
$\alpha_{P} = 2/3$ is the nominal scaling predicted 
by self-similar hydrostatic equilibrium models \citep[A10]{nagai07}.\footnote{
  Unlike e.g. A10 or P12,
  our joint deprojections do not include any
  corrections to the value of $\alpha_{P}$,
  mainly because such corrections
  have negligible effects on our results.
  For example, correcting the value of $\alpha_{P}$ from
  2/3 to our best-fit $\alpha_{P} = 0.49$ found
  in Section~\ref{sec:mass} causes the pressure values in 
  the individual joint deprojection bins for the full sample
  to change by $\lesssim 2$\%.}
We note that our definition of $P_{500}$ is valid for the
electron pressure of the ICM and not the total pressure.

\begin{figure}
  \includegraphics[width=\columnwidth]{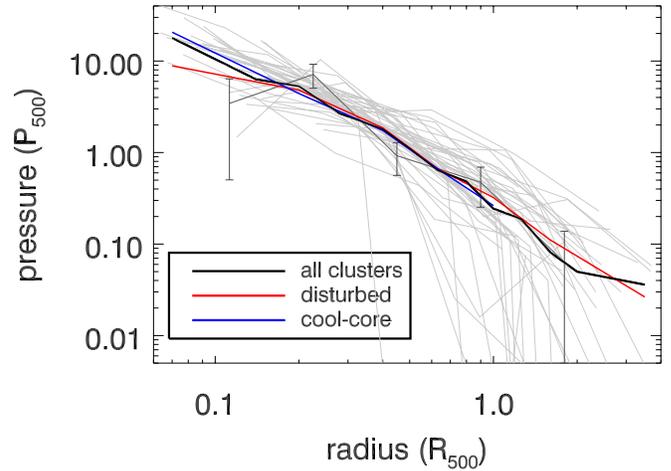}
  \caption{Pressure deprojection for the full BOXSZ
    sample (thick black line), the disturbed subsample (thick red line),
    and the cool-core subsample (thick blue line). 
    Plotted as thin grey lines are the pressure deprojections for 
    each of the 45 clusters in the BOXSZ sample.
    Each individual cluster was deprojected into 5 radial bins located
    at 0.5, 1.0, 2.0, 4.0, and 8.0~arcmin in order to fully 
    sample the spatial dynamic range of our images.
    One of the clusters is shown as a darker grey line with error
    bars at each deprojection bin to indicate the typical
    uncertainty on each of the individual cluster deprojections.
    A significant amount of the variation in the individual cluster
    profiles is due to measurement uncertainty, but we do find
    an additional cluster-to-cluster dispersion that is 
    described in Section~\ref{sec:mass}.
    Note that, in all cases, the 
    correlations between bins are large and varied and must
    be accounted for in any fit or interpretation of the data.}
  \label{fig:deprojection}
\end{figure}

We have determined the average pressure profile
of the full BOXSZ sample (along with the disturbed and cool-core subsamples)
via a simultaneous joint deprojection of multiple clusters.
In this approach, a single deprojected profile in units of $P_{500}$
and $R_{500}$ is constrained by the data from an arbitrary number
of clusters using the GLS algorithm described above.
Pressure deprojections for the full BOXSZ sample, the cool-core subsample,
and the disturbed subsample
are shown in Figure~\ref{fig:deprojection} and numerical
values are given in Table~\ref{tab:deprojection}.
For the deprojection of the full BOXSZ sample, we have used 13 approximately
logarithmically spaced bins between 0.07$R_{500}$ and 3.5$R_{500}$.
These deprojections extend beyond the cluster virial radius (typically near
$2R_{500}$, \citet{umetsu11}),
although only in two deprojection bins, each with
a S/N $\simeq 1.5$.
Due to the smaller number of clusters in the two subsamples, we have used
7 bins spanning the same radial range in each case.
We note that, for many of the clusters in our sample, particularly the
preferentially higher redshift disturbed clusters, the innermost
radial bin(s) is(are) inside of Bolocam's PSF half-maximum radius of 29~arcsec
(in the most extreme case 0.07$R_{500}$ is $\simeq 8$~arcsec).
However, we have shown that even for the disturbed subsample we
are able to obtain unbiased fits to the innermost bins for deprojections
of model input clusters, indicating that our joint deprojections are sensitive
to the shape of the pressure profile at these sub-PSF-sized scales.

\begin{deluxetable*}{cccc} 
  \tablewidth{0pt}
  \tablecaption{Pressure Deprojections}
  \tablehead{\colhead{radius} & \colhead{all clusters} &
    \colhead{disturbed} & \colhead{cool-core} \\
    \colhead{$R_{500}$} & \colhead{$P_{500}$} &
    \colhead{$P_{500}$} & \colhead{$P_{500}$}}
  \startdata
    0.07 & $(1.79 \pm 0.30) \times 10^{+1}$ & 
    $(0.87 \pm 0.16) \times 10^{+1}$ & 
    $(2.07 \pm 0.14) \times 10^{+1}$ \\
    0.14 & $(6.32 \pm 1.48) \times 10^{0\phm{-}}$ & - & - \\
    0.20 & $(5.31 \pm 1.11) \times 10^{0\phm{-}}$ & 
    $(4.81 \pm 0.43) \times 10^{0\phm{-}}$ &
    $(4.41 \pm 0.29) \times 10^{0\phm{-}}$ \\
    0.28 & $(2.68 \pm 0.38) \times 10^{0\phm{-}}$ & - & - \\
    0.40 & $(1.81 \pm 0.23) \times 10^{0\phm{-}}$ & 
    $(1.86 \pm 0.15) \times 10^{0\phm{-}}$ & 
    $(1.74 \pm 0.15) \times 10^{0\phm{-}}$ \\
    0.50 & $(1.11 \pm 0.13) \times 10^{0\phm{-}}$ & - & - \\
    0.63 & $(6.41 \pm 1.01) \times 10^{-1}$ &
    $(6.69 \pm 0.94) \times 10^{-1}$ &
    $(6.64 \pm 0.93) \times 10^{-1}$ \\
    0.80 & $(4.82 \pm 0.69) \times 10^{-1}$ & - & - \\
    1.00 & $(2.44 \pm 0.61) \times 10^{-1}$ &
    $(3.22 \pm 0.62) \times 10^{-1}$ &
    $(2.63 \pm 0.56) \times 10^{-1}$ \\
    1.26 & $(1.87 \pm 0.45) \times 10^{-1}$ & - & - \\
    1.60 & $(0.81 \pm 0.39) \times 10^{-1}$ &
    $(1.11 \pm 0.40) \times 10^{-1}$ &
    $(\le 0.62) \times 10^{-1}$ \\
    2.00 & $(4.99 \pm 3.72) \times 10^{-2}$ & - & - \\
    3.50 & $(3.60 \pm 2.63) \times 10^{-2}$ &
    $(2.64 \pm 2.12) \times 10^{-2}$ &
    $(\le 8.13) \times 10^{-2}$
  \enddata
  \tablecomments{Deprojected pressure profiles for the full BOXSZ sample, 
  the disturbed subsample, and the cool-core subsample.
  The error bars give the square root of the diagonal elements
  of the covariance matrix.
  The best-fit pressure in the two outermost
  bins of the cool-core deprojection is consistent with zero, and 
  so we instead list 68\%
  confidence level upper limits on those values.}
  \label{tab:deprojection}
\end{deluxetable*}

\begin{figure}
  \includegraphics[width=\columnwidth]{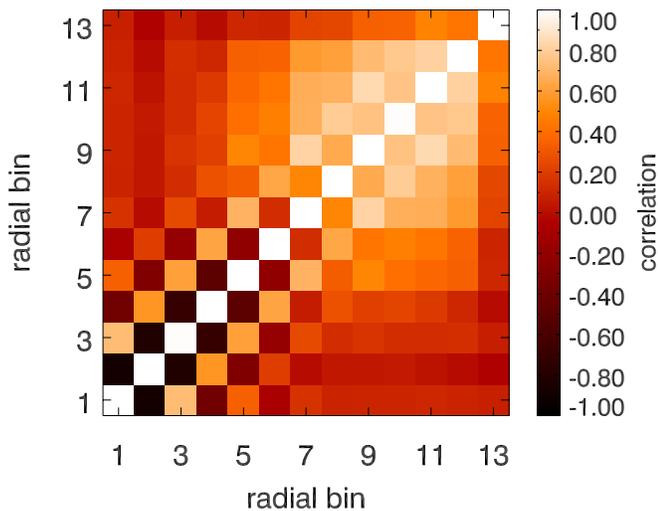}
  \caption{Correlation matrix for the pressure deprojection of the
    full data set into 13 radial bins from 0.07$R_{500}$ to 3.5$R_{500}$
    (see Table~\ref{tab:deprojection} for the exact radius of each bin).
    At large radii the adjacent bins have
    significant positive correlations due to the high pass filtering
    applied to the data, while at small radii the high-pass filtering
    has little effect on the data,
    and consequently there are significant
    anti-correlations between adjacent bins at those radii.}
  \label{fig:correlation}
\end{figure}

We determine the covariance matrix of our deprojected pressure
profiles directly from the set of 1000 deprojections of 
model-plus-noise data. 
The off-diagonal elements of the covariance matrix are in general
non-zero and positive at large radii, indicating significant positive correlations
between most of those deprojection bins (see Figure~\ref{fig:correlation}).
This can be understood as a consequence of the high-pass filtering 
applied to our data, which results in SZ images that are sensitive
to the large-scale shape of the pressure profile 
(i.e., $dP/dR$) but not its absolute normalization.
In contrast, the high-pass filtering does not have a significant effect
on the deprojection bins at small radii, and adjacent bins at small radii tend to 
have strong negative correlations.
These large correlations, which exist for both the jointly and independently 
deprojected profiles, must be accounted for in any interpretation of 
our results.
As we noted above, our estimation of the correlation
matrix assumes a particular cluster pressure profile.
We again tested this dependence by computing a correlation matrix
using both our best-fit profile and the best-fit profile of A10 
as inputs, and we find that the rms difference between the 
recovered elements of the covariance matrices for the two
input profiles is 4.3\%.
This value matches the expected variation due to our finite
number of noise realizations, and therefore indicates that
the our method for estimating the 
correlation matrix is independent of our underlying
cluster pressure profile.

In addition to the joint deprojections described above,
we also deprojected the pressure profiles of each cluster individually
using an identical technique.
However, due to the significantly varied range of 
scaled radii probed for each cluster,
deprojecting all of the clusters at the same set of scaled radii
is not ideal.
Consequently, we have deprojected the individual clusters at fixed angular
radii to ensure that they adequately constrain the profile over the
scales probed by our data (0.5, 1.0, 2.0, 4.0, and 8.0~arcmin).
These individual deprojections are shown in Figure~\ref{fig:deprojection}.

\section{Average Pressure Profiles}
\label{sec:average}

In this section we use the deprojections
from Section~\ref{sec:deproj} to constrain the average properties
of cluster pressure profiles using two different techniques, 
and we briefly describe and compare those techniques here.
In Section~\ref{sec:gNFW} we constrain parametrized 
pressure profiles to the joint deprojected profiles.
Such fits have been performed in a number of 
previous analyses (e.g., A10 and P12),
and these fits therefore enable us to directly 
compare our results to the results of those works.
Since these fits ask a relatively simple question of our data 
(i.e., what is the measurement-noise-weighted average pressure profile),
they also allow us to constrain the average profile of our sample
with optimal signal to noise over a broad radial range,
even for physically interesting subsets of the BOXSZ.
In Section~\ref{sec:mass}, we introduce a new technique for
probing the ensemble behavior of cluster pressure profiles
using a Gaussian process formalism.
For this analysis we use the individual cluster pressure deprojections
to simultaneously constrain an ensemble mean pressure profile,
a covariance matrix describing the intrinsic scatter about
this mean profile, and the mass scaling of the pressure profiles.
This novel technique asks a more demanding set of questions of
our data compared to a parametric fit of the average profile,
and consequently limits us to constraints over a smaller radial range.

\subsection{Parameterized Pressure Profiles}
\label{sec:gNFW}

One method for characterizing the gross behavior of pressure profiles in our sample 
(or a subsample) is to fit a parameterized function to the data.
For this analysis,
we fit parametric models to our deprojected data, rather than directly to
the map data, because fitting directly to the map data takes
an inconveniently large amount of computing time, several days.
As we describe in detail below, we found that parametric fits to
our deprojected profiles, which require only a few minutes,
 were indistinguishable from parametric
fits directly to the map data.
We emphasize that there is no fundamental reason why the fits cannot
be performed directly to the map data, and our choice to fit the
deprojected profiles was motivated entirely by computational
expediency.

To perform the parametric fits to our deprojected data,
we make the simplifying assumption that the noise is Gaussian
and therefore fully described by its covariance matrix.
We applied Mardia's test of multivariate normality to confirm
this assumption is correct \citep{mardia70}.
Mardia's skewness statistic has a limiting distribution equal to the
$\chi^2$ distribution, and we find a $\chi^2$ per degree of 
freedom (DOF) equal to $\chi^2_{\textrm{red}} = 465/455$
(corresponding to a probability to exceed (PTE) of 0.36).
Mardia's kurtosis statistic has a limiting distribution equal to the unit
normal distribution, and we find a value of $-0.96$
(corresponding to a two-sided PTE of 0.34).
Therefore, the noise properties of our deprojected profiles are
consistent with Gaussian distributions.
In addition, we directly fit a single generalized Navarro, Frenk, and 
White \citep[NFW, ][]{navarro96} model to our
map data, allowing all five parameters to vary. This model is described by
\begin{displaymath}
      \tilde{P}(X) = \frac{P_0}{( C_{500} X )^{\mathcal{\gamma}} [1 + 
      ( C_{500} X )^\mathcal{\alpha} ]^{(\mathcal{\beta}
      -\mathcal{\gamma})/\mathcal{\alpha}}},
\end{displaymath}
where $\tilde{P}(X)$ is the scaled pressure profile (in units of $P_{500}$),
$X = R/R_{500}$, $P_0$ is the normalization, $C_{500}$
sets the radial scale, and $\alpha$, $\beta$, and $\gamma$ describe
the power law slope at moderate, large, and small radii \citep{nagai07}.
Similarly to our deprojection procedure, we then determined our parameter
uncertainties using fits to 1000 separate noise realizations.
For comparison, we then performed a Markov chain monte carlo (MCMC) 
fit of the same model to our deprojected profile 
assuming a Gaussian covariance matrix.
We find that the profiles recovered from fitting the
map data directly and from fitting the deprojected data agree within
$\lesssim 1$\% for the range of our deprojected data
($0.07R_{500} \le R \le 3.5R_{500}$),
indicating that the two methods recover consistent results.
Finally, we have verified that there are no biases in
our parametric model fits to either the map data or the deprojected data
by performing both types of fits to a simulated cluster with a known profile.
Specifically, we added a cluster with the best-fit A10 pressure profile to each of the
1000 noise realizations for each cluster, and repeated our analysis
on each of these simulated data sets.
We find that the average best-fit gNFW profiles to these data, both
from direct fits to the map data and from fits to the joint deprojections,
agree with the input A10 profile within $\lesssim 1$\% between
$0.07R_{500}$ and $3.5R_{500}$.
Since we find no significant difference between gNFW fits directly
to the map data and gNFW fits obtained from the deprojected data,
we obtained all of the results described below from fits to the 
deprojected data.

\begin{figure}
  \includegraphics[width=\columnwidth]{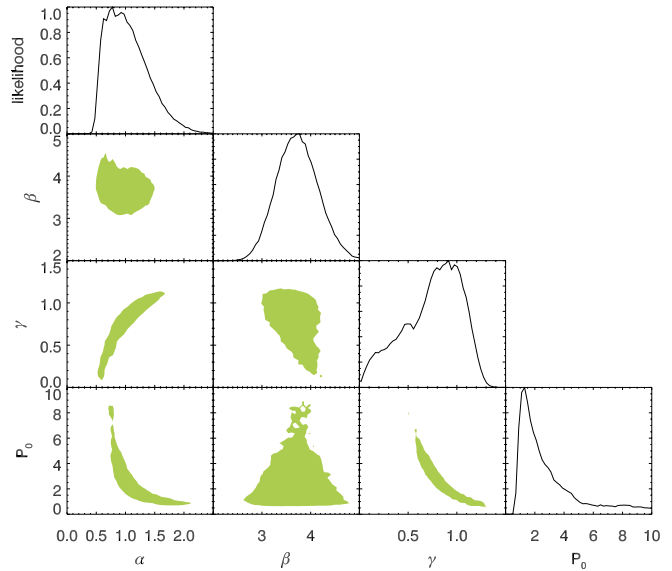}
  \caption{Two-parameter confidence regions (68\%) and one-parameter
    likelihoods for a gNFW fit to our joint deprojected profile
    of the full dataset. From left to right
    and top to bottom the plots show $\alpha$, $\beta$, 
    $\gamma$, and $P_0$ with fixed $C_{500} = 1.18$.
    The large degeneracies between the fit parameters are clearly seen,
    along with the corresponding lack of constraining power on any
    individual parameter.
    However, as shown in Figure~\ref{fig:gnfw_1}, the overall
    pressure profile is tightly constrained.}
  \label{fig:fit_contours}
\end{figure}

We find that our data do not constrain all five of the gNFW fit
parameters within the physically allowed region.
Specifically, our best-fit outer slope of $\beta = 2.03$ implies
an infinite total pressure.\footnote{
  Here, and throughout this manuscript, we present best-fit
  parameter values without error estimates due to the large
  degeneracies between parameters, as illustrated in
  Figure~\ref{fig:fit_contours}.}
As a result, we refit the data fixing the concentration parameter to $C_{500} = 1.18$
(the best fit value found by A10),
obtaining a physically allowed outer slope of $\beta = 3.67$.
The quality of this four-parameter fit is good ($\chi^2_{\textrm{red}} = 1.0$ for 9 DOF),
indicating that this four-parameter gNFW fit describes our data
with a sufficient goodness of fit.
We note that the actual pressure profile over the radial range
constrained by our data 
($0.07R_{500} \lesssim R \lesssim 3.5R_{500}$) for
this four-parameter gNFW fit is only slightly different from the five-parameter fit,
indicating that the unphysical outer slope in the five-parameter gNFW fit
is likely due to the finite radial extent of our data.
Therefore, the best-fit gNFW model of the BOXSZ sample is given by
\begin{displaymath}
[C_{500}, \alpha, \beta, \gamma, P_0] = 
  [{\it 1.18}, 0.86, 3.67, 0.67, 4.29],
\end{displaymath}
where the value of $C_{500}$ is in italics to emphasize that it was held
fixed at 1.18 in our fits.
Our choice to fix the value of $C_{500}$, rather than one of the other
parameters, was motivated by the fact that all of the possible four-parameter
gNFW fits have similar fit qualities, but any combination that varies
both $C_{500}$ and $\beta$ results in an unphysical outer slope.
The set of two-dimensional confidence regions for the four-parameter gNFW fit
is illustrated in Figure~\ref{fig:fit_contours}, and the strong
parameter degeneracies mentioned above are evident in these plots.
To test for possible biases associated with 
our choice to fix $C_{500}$ to
the best-fit value determined by A10, we also computed gNFW fits with the value
of $C_{500}$ set to the best-fit value of P12, $C_{500} = 1.81$, and to the best-fit
value of \citet{nagai07}, $C_{500} = 1.30$,\footnote{
  In their unpublished erratum, \citet{nagai07} find $C_{500} = 1.3$ rather
  than the published value of $C_{500} = 1.8$.}
with the results shown in Table~\ref{tab:gnfw_fits}.
Compared to our choice of $C_{500} = 1.18$,
we find that these values of $C_{500}$ result in similar fit qualities 
and profiles that differ by an rms of 4\% (P12) and 1\% \citep{nagai07}
over the radial range $0.07R_{500} \le R \le 3.5R_{500}$.
Therefore, 
we conclude that our choice to fix $C_{500}$ to the best-fit value determined 
by A10 has little impact on our results.

\begin{deluxetable*}{ccccccccc} 
  \tablewidth{0pt}
   \tablecaption{gNFW Fits to the Bolocam Cluster Sample}
   \tablehead{\colhead{$C_{500}$} & \colhead{$\alpha$} &
     \colhead{$\beta$} & \colhead{$\gamma$} &
     \colhead{$P_{0}$ $(P_{500})$} & \colhead{$\chi^2_{\textrm{red}}$} & 
     \colhead{DOF} & \colhead{PTE} & \colhead{notes}}
   \startdata
\sidehead{all clusters}
   1.81$^{\phm{\star}}$ & 1.33$^{\phm{\star}}$ & 4.13$^{\phm{\star}}$ & 0.31$^{\phm{\star}}$ & 6.54$^{\star}$ & 4.7 & 12 & 0.00 & best fit shape from P10 \\
   1.18$^{\phm{\star}}$ & 1.05$^{\phm{\star}}$ & 5.49$^{\phm{\star}}$ & 0.31$^{\phm{\star}}$ & 7.82$^{\star}$ & 2.7 & 12 & 0.00 & best fit shape from A10 \\
   \textbf{1.18$^{\phm{\star}}$} & \textbf{0.86$^{\star}$} & \textbf{3.67$^{\star}$} & \textbf{0.67$^{\star}$} & \textbf{4.29$^{\star}$} & 1.0 & 9 & 0.44 & best fit from this work (using A10 value of $C_{500}$) \\ 
   1.81$^{\phm{\star}}$ & 1.32$^{\star}$ & 2.91$^{\star}$ & 0.92$^{\star}$ & 2.60$^{\star}$ & 1.0 & 9 & 0.48 & 4-parameter fit using P12 value of $C_{500}$ \\
   1.30$^{\phm{\star}}$ & 0.91$^{\star}$ & 3.51$^{\star}$ & 0.71$^{\star}$ & 3.94$^{\star}$ & 1.0 & 9 & 0.45 & 4-parameter fit using \citet{nagai07} value of $C_{500}$ \\
   3.19$^{\star}$ & 3.28$^{\star}$ & 2.03$^{\star}$ & 1.10$^{\star}$ & 3.07$^{\star}$ & 0.9 & 8 & 0.52 & 5-parameter fit, yields an unphysical outer slope $\beta$ \\
\sidehead{disturbed clusters}
   \textbf{1.18$^{\phm{\star}}$} & \textbf{0.90$^{\star}$} & \textbf{5.22$^{\star}$} & \textbf{0.02$^{\star}$} & \textbf{17.28$^{\star}$} & 3.1 & 4 & 0.02 & best fit from this work \\ 
\sidehead{cool-core clusters}
   \textbf{1.18$^{\phm{\star}}$} & \textbf{2.79$^{\star}$} & \textbf{3.51$^{\star}$} & \textbf{1.37$^{\star}$} & \textbf{0.65$^{\star}$} & 1.7 & 4 & 0.15 & best fit from this work \\ 
   \enddata
   \tablecomments{gNFW fits to the deprojected profiles computed for the BOXSZ sample. 
   From left to right the columns give the concentration parameter relative to $R_{500}$
   ($C_{500}$), the power law slopes at intermediate, large, and small radii ($\alpha, \beta, \gamma$),
   the normalization $P_0$ relative to $P_{500}$, the reduced $\chi^2$ of the fit,
   the degrees of freedom in the fit, the associated PTE, and any notes regarding the particular fit.
   The upper rows show fits to the deprojection of the full sample, and show parameter values
   and fit qualities when different numbers of parameters are varied.
   The lower two rows show the best-fit four-parameter models for the disturbed
   and cool-core subsamples.
   In all cases, stars denotes parameters that were varied in the fit.
   In the first two rows, we fit the models of P12 and A10, 
   varying only the normalization.
   In all subsequent rows, the fixed parameters
   were set to the values found in A10.
   We find that varying four parameters provides a sufficient goodness of fit to the data, which
   is only marginally improved when all five parameters are varied.
   However, varying all five parameters results in a profile with an unphysically
   small outer slope ($\beta=2.03$), and we therefore take the four-parameter
   fit as the best description of our data.}
   \label{tab:gnfw_fits}
\end{deluxetable*}

\begin{figure}
  \includegraphics[width=\columnwidth]{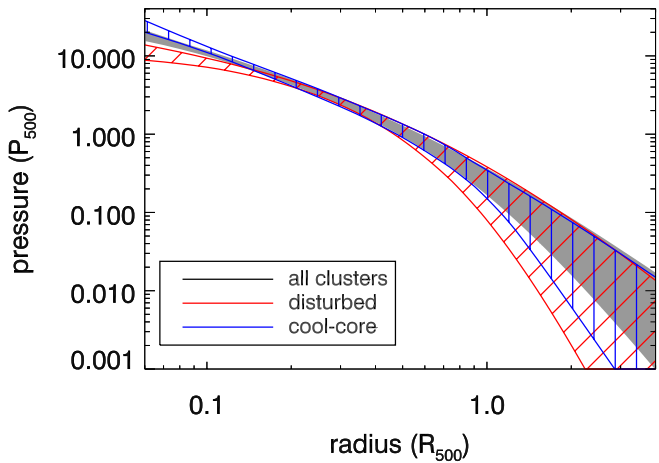}
  \caption{gNFW parameterized fits to the Bolocam data, varying 
    four parameters of the gNFW model ($C_{500}$ was fixed to 1.18).
    The bands indicate the maximum and minimum pressure values
    as a function of radius, bounding 68\% of the MCMC fits.
    The plot shows fits to the full BOXSZ, the
    disturbed subsample, and the cool-core subsample.
    The pressure profile appears to be independent of cluster
    morphology at $R \gtrsim 0.15R_{500}$,
    but the cool-core clusters have higher
    pressures than the disturbed clusters at smaller
    radii.}
  \label{fig:gnfw_1}
\end{figure}

We then fit the same four-parameter gNFW model to both
the disturbed and cool-core subsamples of the BOXSZ
cluster sample (see Table~\ref{tab:gnfw_fits}).
In good agreement with previous results from A10 and P12,
we find consistent average pressure profiles, given our measurement 
uncertainties, between these subsamples
(and the full sample) at all intermediate and large
radii ($R \gtrsim 0.15R_{500}$).
At smaller radii, the profiles clearly diverge, and 
at those radii the cool-core
clusters have a higher pressure than the
disturbed clusters (see Figure~\ref{fig:gnfw_1}).\footnote{
  As we described in detail in Section~\ref{sec:sample},
  there is an asymmetry in the redshift distributions
  of the cool-core and disturbed subsamples.
  Consequently, we cannot definitively rule out
  redshift evolution as the cause of the
  discrepancy in the pressure profiles at small radii.
  However, given that A10 observed a similar difference
  in the inner pressure profiles of disturbed and cool-core
  systems for a sample of low-redshift clusters,
  the differences we observe are likely due to morphology
  rather than redshift evolution.}
Although the overall pressure profiles for the full BOXSZ sample,
the cool-core subsample, and the disturbed subsample are quite similar
at most radii, we note that the best-fit gNFW parameters are quite different,
further emphasizing the large degeneracies between these parameters.
Furthermore, the fit quality of the gNFW model is slightly worse
for both of the two subsamples compared to the full sample,
with PTEs of 0.02 and 0.15 for the disturbed and 
cool-core fits compared to a PTE of 0.44 for the full-sample fit.

\begin{figure*}
  \includegraphics[width=\textwidth]{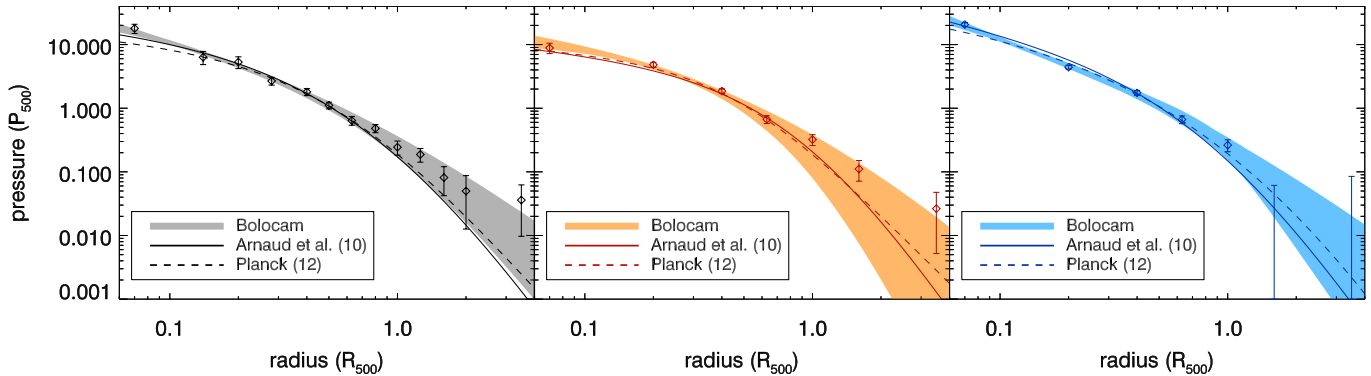}
  \caption{gNFW parameterized fits to the BOXSZ sample, varying
    four parameters of the gNFW model.
    From left to right the three plots show the 
    Bolocam fit to the full
    sample, disturbed subsample, and cool-core
    subsample as points with error bars and with 
    shaded regions representing the 68.3\% confidence region
    for the gNFW fits
    (We note that gNFW fits with five free parameters
    are slightly more consistent with the measured data
    at large radius, but, as described in the text,
    these fits were discarded because they produce
    unphysical outer slope).
    The best-fit parameterizations given
    in A10 and P12 are overlaid as thin and dashed lines
    (P12 did not fit a disturbed subsample, so we overlay
    their non-cool-core fit in the center plot).
    The A10 fits relied on the REXCESS sample of 33 low-$z$ clusters ($z < 0.2$)
    observed with \emph{XMM-Newton} within $R_{500}$ and 
    results from simulations outside $R_{500}$. 
    The P12 fits relied on a sample of 62 \emph{Planck} selected
    clusters at $\langle z \rangle \simeq 0.15$, and used \emph{XMM-Newton} data
    to constrain the inner portion of the profile and \emph{Planck}
    data to constrain the outer portion of the profile.
    Our fits use Bolocam SZ data for a sample of 45 higher redshift clusters
    ($0.15 \le z \le 0.89$).}
  \label{fig:gnfw_2}
\end{figure*}

\begin{figure*}
  \includegraphics[width=.45\textwidth]{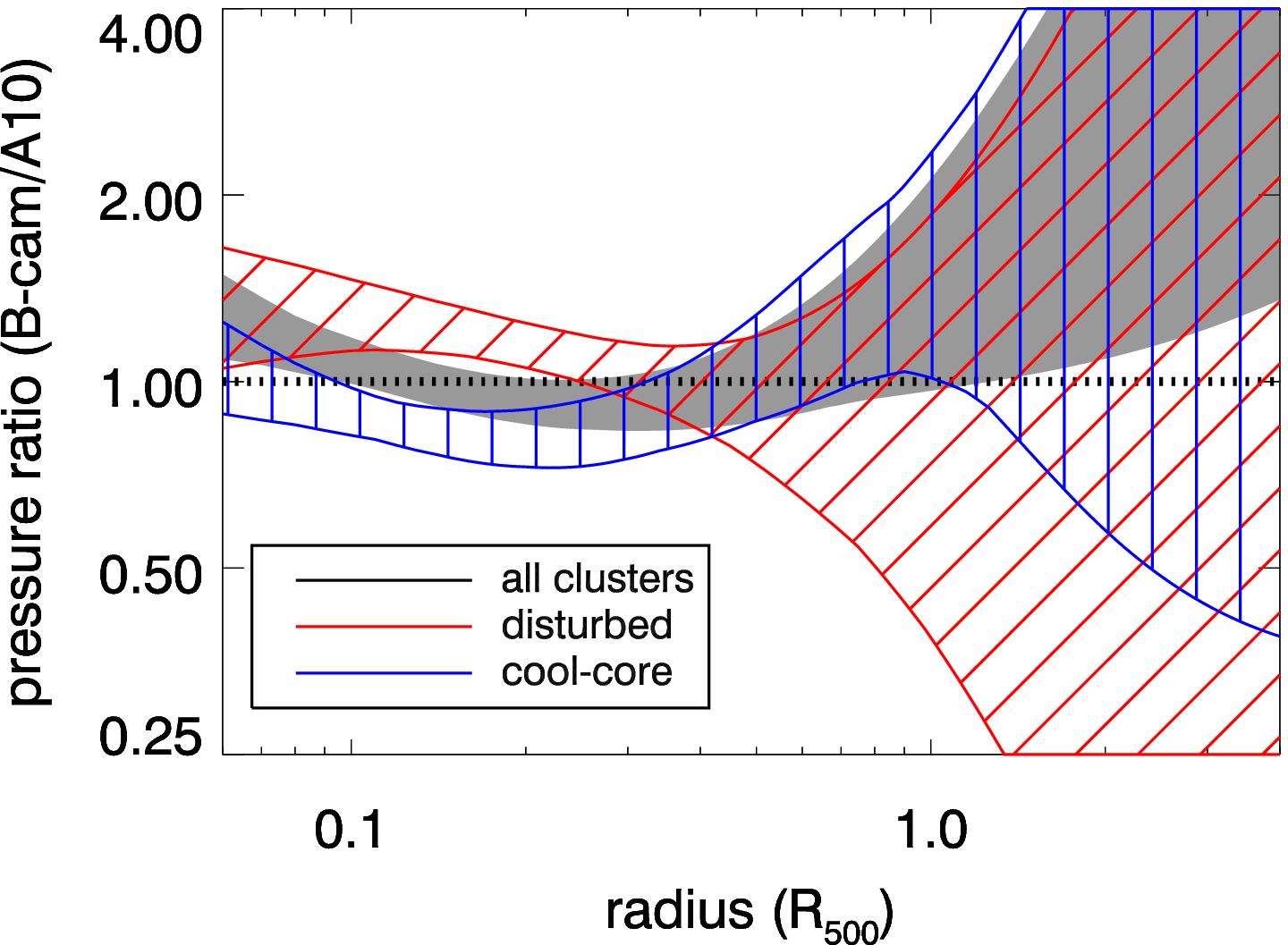}
  \hspace{0.095\textwidth}
  \includegraphics[width=.45\textwidth]{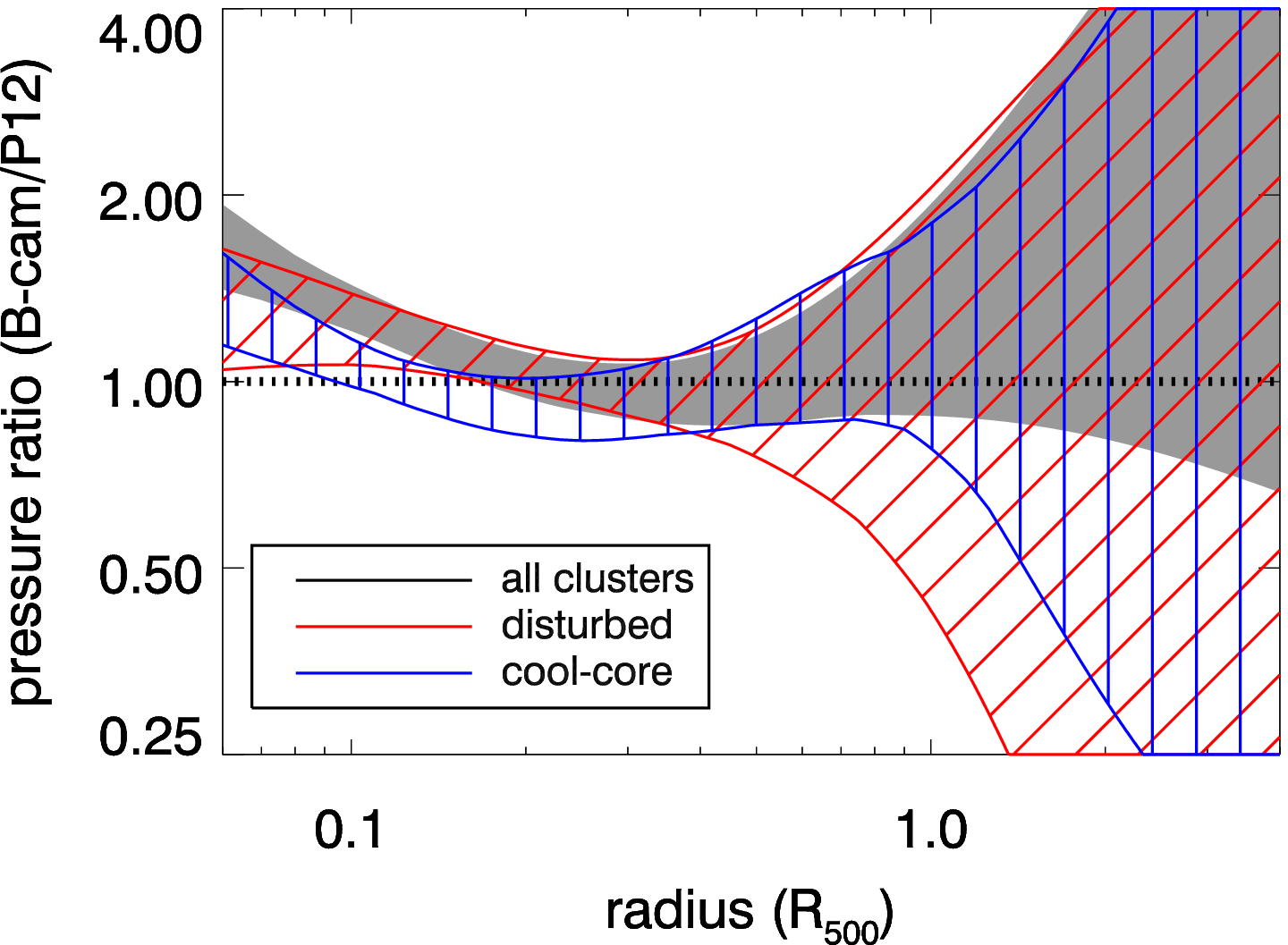}
  \caption{Confidences regions (68.3\%) for
    the ratio of our best-fit four-parameter gNFW
    fits to the best-fit four-parameter gNFW fits from
    A10 (left) and P12 (right).
    In both cases the agreement is generally good in
    the regions that are well constrained by all 
    three datasets 
    ($0.1R_{500} \lesssim R \lesssim 1.0R_{500}$).
    However, the fit to the full BOXSZ sample shows hints of higher
    pressure than the A10 fit at both large and small radii,
    and hints of higher pressure than the P12
    fit at small radius.}
  \label{fig:gnfw_3}
\end{figure*}

Due to the large degeneracies between the parameters in the
gNFW model, it is difficult to quantify the differences
between our best-fit gNFW model and those found in previous
analyses via a direct comparison of the fit parameters
(\citealt{nagai07}, A10, \citealt{plagge10}, P12).
Consequently, we have compared the profiles resulting from
our gNFW fits to the gNFW profiles found by
A10 and P12 over the
approximate radial range constrained by all three datasets
(0.05$R_{500} \lesssim R \lesssim 4R_{500}$, see
Figures~\ref{fig:gnfw_2} and \ref{fig:gnfw_3}).
We in general find excellent agreement between our pressure
profiles and those found in these previous analyses, 
regardless of morphological classification\footnote{
  P12 only present results for cool-core
  and non-cool-core subsamples, and we therefore 
  take their non-cool-core results to be representative
  of disturbed systems.}
(e.g., the cool-core profile from our analysis is in good
agreement with the cool-core profile of A10).
We do note that our disturbed and cool-core systems indicate
slight differences within 0.3$R_{500}$ compared to the
corresponding results of A10, and our overall average
profile indicates slightly 
higher pressures at $R \lesssim 0.1R_{500}$
and at $R \gtrsim 1.0R_{500}$ compared to 
the results of A10.
Our overall average profile also shows higher pressure at
small radii compared to the results of P12.
Our results are therefore more similar to
simulation-derived results at small radii, which also
show higher pressures than found by A10 and P12 
\citep[A10, ][P12]{borgani04, nagai07, piffaretti08}.

The overall good agreement between our best-fit gNFW profiles
and the results derived in previous analyses provides
further evidence that the average cluster pressure profile is
approximately universal (at least within our measurement uncertainties
on the average profile, which are $\simeq 10$--$20$\% inside $R_{500}$).
This is especially true given the large differences in
the median redshifts ($\langle z \rangle = 0.12$, 0.15, and 0.42),
median masses ($\langle M_{500} \rangle = 3$, 6, and $9 \times 10^{14}$~M$_{\odot}$)
and data types (X-ray/simulation, X-ray/SZ, SZ-only)
for the A10, P12, BOXSZ samples.
As another consistency check between our data and the results
of A10 and P12, we fit each of
their best-fit gNFW models to our data, allowing only the normalization
to be a free parameter.
Although the fit quality is poor in both cases, we 
find normalizations consistent with both results
(7.82 compared to 8.40 for the A10 fit and
6.54 compared to 6.41 for the P12 fit)\footnote{
  Although these single-parameter gNFW fits do not suffer
  from the same degeneracies seen in the multi-parameter
  fits, we forgo error estimates because the poor fit
  quality calls into question the accuracy of such estimates.
  However, the fits indicate that our uncertainty on
  the normalization is likely to be dominated by
  our 5\% flux calibration uncertainty.
  Since our values differ from the A10 and
  P10 values by 7\% and 2\%, we can conclude that the
  normalizations found by all three datasets are consistent.}.
This implies that the average total pressure of our sample
is consistent with the average total pressure found in those
analyses, further showing the approximate universality
of cluster pressure profiles and the good agreement
between SZ and X-ray measurements of those profiles.
In addition, we note that two other analyses show good
agreement with the results of A10 (and consequently our 
results as well).
\citet{sun11} analyzed {\it Chandra} data for a set of 43 low redshift groups
and found an average  pressure profile that is within 1--$\sigma$ of the A10 profile
over nearly all of its range ($0.01R_{500} \lesssim R \lesssim R_{500}$).
\citet{plagge10} used SPT 
SZ data for a set of 15 moderate redshift clusters to constrain a gNFW
profile, finding shape parameters that are statistically consistent
with the A10 values (3/4 paramters agree within 1--$\sigma$,
and the fourth agrees within 2--$\sigma$).

In addition to fitting gNFW models to our data, we also
fit a $\beta$-model of the form
\begin{displaymath}
  \tilde{P}(X) = \frac{P_0}{\left( 1 + ( X/R_c )^2 \right)^{3 \beta/2}},
\end{displaymath}
where $P_0$ is the pressure normalization, $R_c$ is the core
radius, and $\beta$ is the power law slope
\citep{cavaliere76, cavaliere78}.
Fitting the full BOXSZ sample, we find best-fit parameters of
$R_c = 0.11$ (relative to $R_{500}$),
$\beta = 0.61$ and $P_0 = 18.9$. 
We find $\chi_{\textrm{red}}^2 = 1.5$ for 10 DOF,
which gives a PTE of 0.13 and indicates a somewhat worse fit
compared to the gNFW model.
The fit appears to be largely driven by the higher S/N data at small radii,
which explains why the best-fit value of $\beta$ is more
similar to X-ray derived results from fits to the inner regions of clusters
($\beta \simeq 2/3$, e.g., \citealt{jones84, arnaud09}) than those 
derived from X-ray surface brightness profiles at large radii
(e.g., \citealt{vikhlinin99, maughan08})
or from SZ data extending to large radius 
($\beta \simeq 0.85-1.05$, e.g., \citealt{hallman07, plagge10}).
Given the superior fit quality of the gNFW model, along with
the known shortcomings of the $\beta$-model (e.g., \citealt{mohr99, hallman07}), 
we do not explore the $\beta$-model in any additional detail.

\subsection{Gaussian Process Description of the Ensemble Properties of the Pressure Profiles}
\label{sec:mass}

Jointly fitting distinct subsets of clusters provides some information on the differences among cluster 
pressure profiles as a function of radius, but does not directly probe the intrinsic scatter
among these profiles. 
Investigating the scatter instead requires a model for the ensemble of profiles to be fit to the 
data from individual clusters. Here we adopt arguably the simplest such model, describing 
scaled cluster pressure profiles as a Gaussian process (for an introduction to Gaussian processes, see e.g. 
\citealt{rasmussen2005gaussian}). 

In this approach, the ensemble of profiles is modeled by (1) a mean scaled pressure 
profile as a function of scaled radius, $\bar{P}(x)$,\footnote{
  Note that this ensemble mean profile is conceptually different from the average parameterized 
  profiles fit in Section~\ref{sec:gNFW}. The former describes the average profile accounting for 
  the presence of intrinsic scatter, i.e. the center of an ensemble of profiles at a given radius, 
  while the latter assumes that all cluster profiles are described by a single function, 
  with residuals between the model and data entirely due to known measurement uncertainties.} 
and (2) a covariance function, 
$\Sigma(x,y)$, encoding the intrinsic scatter about $\bar{P}(x)$ as a function of radius.
Mathematically, the likelihood for a single realization of $P(x)$ to have a set of scaled 
pressures $\{P_i\}$ at scaled radii $\{X_i\}$ is proportional to
\[
  \frac{\exp\left( -\frac{1}{2}z^\mathrm{T} S^{-1} z \right)}{|S|^{1/2}},
\]
where $z_i = P_i - \bar{P}(X_i)$ and $S_{i,j} = \Sigma(X_i,X_j)$, for any 
pair of radial values $X_i$ and $X_j$.
Thus, the diagonal covariance terms, 
$\Sigma(x,x)$ dictate the marginal intrinsic scatter 
among profiles at a given radius; while off-diagonal terms, $\Sigma(x,y)$ with $x \neq y$,
determine whether realizations of the profile tend to be shifted coherently with respect to 
$\bar{P}$ (positive values), or tend to cross $\bar{P}$ (negative values).
In addition to the mean profile and scatter, we fit simultaneously for the mass dependence of the 
pressure normalization via the parameter $\alpha_{P}$ in Equation~\ref{eqn:p500}.\footnote{
    Including a free power of $E(z)$ has a negligible effect on our 
    results in this section. Since the data constrain this evolution term very poorly, 
    we did not investigate such an evolution term further.}

\begin{figure*}
  \includegraphics[width=0.45\textwidth]{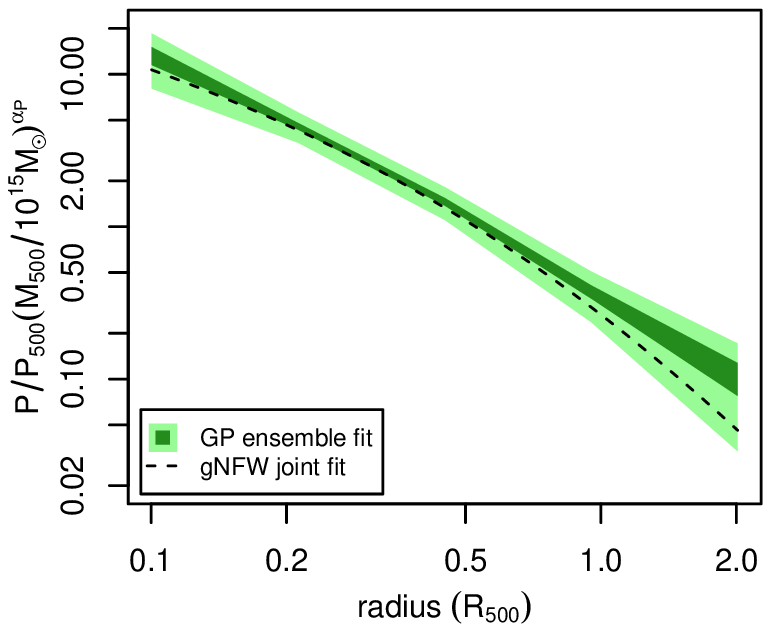}
  \hspace{.095\textwidth}
  \includegraphics[width=0.45\textwidth]{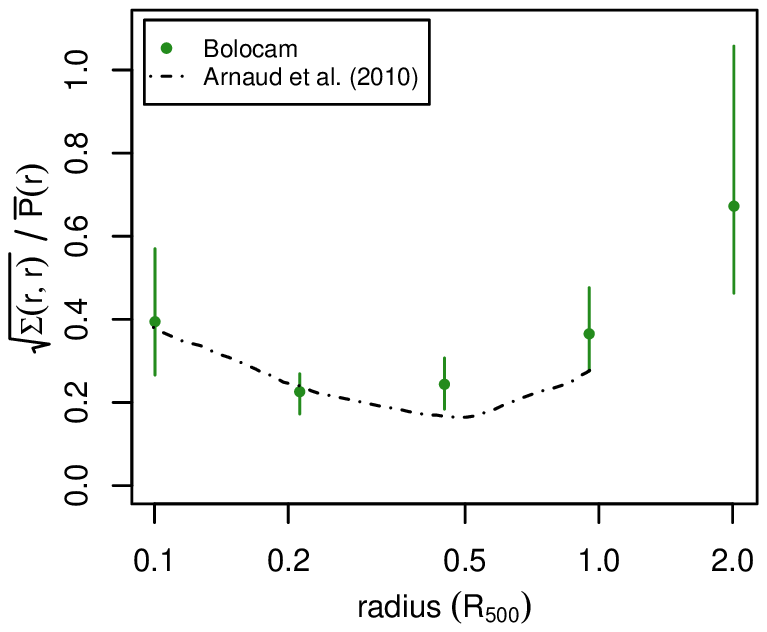}
  \caption{Left: the dark, inner shaded region shows the 68.3\% confidence posterior for 
    the mean pressure profile determined from our Gaussian process analysis, 
    while the light, outer region indicates the 
    best-fit marginal intrinsic scatter at each radius (the square root of 
    the diagonal elements of the covariance matrix). At all radii, the 
    uncertainty on the mean function is smaller than the corresponding 
    nominal intrinsic scatter. The dashed line shows the best gNFW joint fit, which 
    assumes that a single profile describes all clusters, and differs slightly from the 
    Gaussian process fit, which includes intrinsic scatter.
    Right: the fractional intrinsic scatter (diagonal covariances scaled by 
    the best fitting mean profile) as a function of radius.
    At all radii, zero intrinsic scatter is excluded at $> 95.4$\% confidence. 
    The intrinsic scatter estimated by A10 from X-ray data is shown as the dot-dash line, 
    and is in good agreement with our results at $R<R_{500}$.
  }
  \label{fig:GP}
\end{figure*}

The individually deprojected cluster profiles from Section~\ref{sec:deproj} will be used to 
constrain this model. However, those profiles are each constrained at different scaled radii, 
presenting a significant complication to the analysis. To simplify the problem, we interpolate 
the individual profiles to a set of 5 common scaled radii, logarithmically spaced between 
$0.1R_{500}$ and $2.0R_{500}$. The radial range probed here is smaller than that covered 
by the sample as a whole, reflecting the fact that data from a sufficient number of clusters 
must exist at each radius in order to constrain the scatter. 
The interpolation was accomplished as follows.
First, we generate a multivariate normal draw of the pressures
at the deprojection radii using the measured mean values and measurement error
covariance matrices of the individual cluster deprojections from Section~\ref{sec:deproj}.
We then interpolate these pressures to the set of 5 common
scaled radii using a power-law interpolation (recall that our deprojections
assumed power-law behavior between the deprojection radii).
This process is repeated many times in order to constrain the mean
pressures and the measurement error covariance matrices at the common scaled radii.
Since the individual profiles do not cover the entire range $0.1<X<2$, each cluster 
provides information at only a subset of the final radii.
The result of this procedure is 
that the mean profile and covariance function can be compactly parametrized by 5 pressure values 
(for $\bar{P}$) and the independent elements of a $5\times5$ covariance matrix (for $\Sigma$), 
corresponding to the common scaled radii, where otherwise we would have been forced to assume a 
particular functional form for $\Sigma$. 

Thus, the complete log-likelihood used to constrain the model is
\begin{equation}
  \ln\mathcal{L} = \sum_j -\frac{1}{2} \left[ z_j^\mathrm{T} \left(S+U_j\right)^{-1} z_j + 
    \ln\left|S+U_j\right|\right],
\end{equation}
where the sum is over clusters, and $U_j$ is the measurement error covariance matrix for the 
interpolated, scaled pressure profile of the $j$th cluster. 
The parameter space for this model was explored using MCMC, adopting flat priors on all 
21 free parameters. Maximum-likelihood confidence intervals for each parameter are displayed 
in Table~\ref{tab:GP}, and Figure~\ref{fig:GP} shows the recovered pressure profile
and fractional scatter as a function of radius. The reduced $\chi^2$ of our data with respect 
to the best fitting model is 1.02 for 137 DOF, 
indicating that the Gaussian process description of the 
ensemble of profiles provides a sufficient goodness of fit. 
In particular, while the difference in gNFW profile fits of cool-core and
disturbed clusters is evident at $0.1R_{500}$ in Figure~\ref{fig:gnfw_1},
its modest statistical significance at that radius is reflected in this analysis, 
where a simple Gaussian scatter is seen to be a satisfactory model
for our data.\footnote{
  We also attempted to constrain the intrinsic scatter within the disturbed and cool-core subsamples.
  Unfortunately, neither subsample provides enough data 
  at these small radii for us to constrain ensemble models for the individual subsamples.}

We constrain the mass scaling from Equation~\ref{eqn:p500}
to be $\alpha_{P} = 0.49^{+0.10}_{-0.08}$, which is shallower than the value of 2/3
predicted by self-similar hydrostatic equilibrium scalings \citep[][A10]{kaiser86}.
Under the assumption that the pressure profile shape is independent of mass,
the integrated SZ signal $Y$, or its X-ray analog $Y_X$,
scale with mass according to a power law slope $\alpha_{Y} = \alpha_{P} + 1$.
Our value of $\alpha_{P}$ therefore implies a 
$Y$--$M$ scaling with a power law slope of 1.49, in good agreement with the $Y_X$--$M$ 
power law slope of $1.48\pm0.04$ found by M10 using an identical method for constraining
cluster masses from {\it Chandra} X-ray data.\footnote{
  A separate analysis of the BOXSZ sample, which
  accounts for selection effects and uses directly integrated
  cylindrical Y values within $R_{2500}$, also finds a similar
  slope for the scaling of $Y$ versus $M$ \citep{czakon12}.}
These results are inconsistent with those of A10, who measured a $Y_X$--$M$
scaling with $\alpha_{Y} = 1.78 \pm 0.06$, implying $\alpha_{P} = 0.78$.
We speculate that this discrepancy 
in values of $\alpha_P$ is due to 
differences in mass estimation between A10 and our work (which uses masses determined according to M10). 
Indeed, \citet{rozo12_i,rozo12_ii} have noted that the difference in $Y_X$--$M$ 
slopes between A10 and M10 is consistent with being due to a systematic disagreement in mass estimates, 
after accounting for instrumental calibration. 
Comparable disagreements in X-ray temperature--mass relation slopes, loosely correlated with hydrostatic 
mass calibration techniques, were also pointed out by \citet{mantz11}. 
In addition, as described in Section~\ref{sec:deproj}, we emphasize that
our results for the shape of
the average pressure profile are approximately independent of the exact mass scaling,
and therefore should not be affected by any of these discrepancies in
mass determinations.

Our fit detects the presence of non-zero intrinsic scatter among profiles (diagonal covariance terms) 
at a significant level ($>95.4$\% confidence) at all radii, with the fractional intrinsic scatter 
minimized at radii $\simeq 0.2R_{500}$--$0.5R_{500}$. The off-diagonal covariance terms become consistent with 
zero at large radial separations (i.e. large $|X_1-X_2|$), indicating that our pressure scaling 
has effectively removed the principal mass and redshift dependence of the profiles. 
The adequacy of our simple scaling over a wide redshift and mass range further confirms that cluster 
pressure profiles are approximately universal, with fractional intrinsic scatter away from the 
universal profile at the $\simeq 20$--$40$\% level.

\begin{deluxetable*}{cccccc} 
  \tablewidth{0pt}
   \tablecaption{Gaussian process fit parameters}
   \tablehead{\colhead{radius ($R_{500}$)} & \colhead{0.100} &
     \colhead{0.212} & \colhead{0.449} &
     \colhead{0.951} & \colhead{2.013}}
   \startdata
\sidehead{mean scaled pressure profile, $\bar{P}(X)$} 
 & $  13.3^{+  1.8}_{-  1.8}$ & $  4.59^{+ 0.20}_{- 0.26}$ & $  1.47^{+ 0.08}_{- 0.09}$ & $  0.37^{+ 0.04}_{- 0.04}$ & $  0.102^{+ 0.025}_{- 0.025}$ \\ 
\sidehead{intrinsic scatter covariance matrix, $\Sigma(X_i,X_j)$} 
\vspace{7pt}
0.100 & $  27.5^{+  30.0}_{-  15.0}$ & $ 2.75^{+ 2.50}_{- 2.00}$ & $  0.85^{+ 0.90}_{- 0.80}$ & $  0.22^{+ 0.45}_{- 0.35}$ & $  0.07^{+ 0.26}_{- 0.24}$ \\ 
\vspace{7pt}
0.212 &  & $  1.08^{+ 0.45}_{- 0.45}$ & $  0.19^{+ 0.17}_{- 0.11}$ & $  0.04^{+ 0.07}_{- 0.04}$ & $  0.02^{+ 0.04}_{- 0.04}$ \\ 
\vspace{7pt}
0.449 &  & & $  0.13^{+ 0.08}_{- 0.05}$ & $ 0.037^{+0.024}_{-0.018}$ & $ 0.009^{+0.018}_{-0.012}$ \\ 
\vspace{7pt}
0.951 &  & & & $ 0.019^{+0.013}_{-0.008}$ & $ 0.0038^{+0.0065}_{-0.0055}$ \\ 
\vspace{7pt}
2.013 &  & & & & $0.0047^{+0.0070}_{-0.0025}$ \\ 
   \enddata
   \tablecomments{Gaussian process fit parameters describing the mean pressure profile
     of our sample and the intrinsic scatter about this profile. The fit also
     includes an overall mass scaling, constrained to $\alpha_{P}=0.49^{+0.10}_{-0.08}$; 
     see the text in Section~\ref{sec:mass}.
     All of the diagonal elements of the covariance matrix are different
     from zero at greater than 95.4\% significance, indicating that
     we detect non-zero intrinsic scatter at all radii.
     Most of the off-diagonal elements are consistent with zero,
     indicating that our mass and redshift scaling largely accounts
     for any evolution in the normalization of the pressure profiles.}
   \label{tab:GP}
\end{deluxetable*}
 
In addition, given our measurement uncertainties,
the intrinsic scatter of our high mass and moderate redshift BOXSZ sample is 
consistent with that found using X-ray data within $R_{500}$ for the lower mass and lower 
redshift clusters in the REXCESS sample (A10; Figure~\ref{fig:GP}) 
and for low redshift groups \citep{sun11}, 
supporting the proposition that the intrinsic scatter among pressure profiles is relatively 
independent of mass and redshift. Given the large redshift range spanned by the BOXSZ sample, 
this comparison to results at lower redshifts also indicates that the inexact redshift scalings 
of the \citet{kaiser86} relations do not introduce a significant amount of intrinsic scatter 
among pressure profiles \citep{kravtsov12}. Furthermore, similar to the findings of P12, 
our results imply that the intrinsic scatter continues to increase at radii larger than those 
probed previously by X-ray analyses ($\gtrsim R_{500}$).

We note that the Gaussian process approach employed in this section
provides a compact framework to obtain simultaneous constraints on the 
ensemble mean profile, its scaling relations, and its radius-dependent intrinsic scatter. 
In the present analysis, we have made some assumptions for computational expediency, namely 
that the measurement errors and intrinsic scatter are described well as Gaussian; however, 
neither of these assumptions is necessarily required. Looking forward, such probabilistic 
models in general are attractive because they allow straightforward and quantitative 
comparisons of clusters from different observed samples or from simulations. 
For example, given such a description of simulated cluster profiles, it would be 
straightforward to test whether observed profiles are consistent with the simulations 
using simple $\chi^2$ statistics.

\section{Summary}
\label{sec:summary}

We have examined the pressure profiles determined from Bolocam
SZ observations of the 45 clusters in the BOXSZ sample.
This sample spans a large range in redshift ($0.15 \le z \le 0.89$),
with a median redshift of $z = 0.42$.
These clusters are also among the most massive known, with a median
mass of $M_{500} = 9 \times 10^{14}$~M$_{\odot}$.
All of these clusters have {\it Chandra} X-ray observations,
and we have used these X-ray data to determine the 
mass of each cluster.
Using these masses, we have scaled each SZ pressure profile by
the mass-and-redshift-dependent normalization factor $P_{500}$
and by the overdensity radius $R_{500}$.
We constrained the average pressure profile of the BOXSZ sample
using a joint deprojection technique, using 13 radial bins
approximately logarithmically spaced between 0.07$R_{500}$ and
3.5$R_{500}$.
We note that, since the cluster virial radius is generally near 
$2R_{500}$ \citep{umetsu11},
our deprojected pressure profiles extend beyond
the virial radius, although only in two deprojection bins each with
a S/N $\simeq 1.5$.

The X-ray data were also used to classify the disturbed and cool-core
subsamples of the BOXSZ, and we deprojected the average pressure
profiles of these two subsamples into 7 radial bins spanning
the same radial range (0.07$R_{500}$ to 3.5$R_{500}$).
We fit gNFW models to all three of these average deprojected
pressure profiles, and we find that this model
describes our data with a sufficient goodness of fit
when 4/5 of the gNFW parameters are allowed to vary.
The best-fit average pressure profile of our full sample is described 
by the parameters ($C_{500}$, $\alpha$, $\beta$, $\gamma$, $P_0$ = 
1.18, 0.86, 3.67, 0.67, 4.29).
We find a worse, although acceptable, fit quality
using a $\beta$-model, but we do not explore an in-depth analysis
of $\beta$-model fits due to the known shortcomings of that
model \citep{mohr99, hallman07}.
The gNFW fits show consistent pressure profiles regardless of 
cluster morphological classification outside of $\simeq 0.15R_{500}$
given our measurement uncertainties,
but inside that radius the cool-core systems show 
higher pressures than the disturbed systems.
Due to the large parameter degeneracies in the gNFW model, our
best-fit parameter values are not in general similar to those 
found in previous analyses (\citealt{nagai07, plagge10}, A10, \citealt{sun11}, P12). 
However, the actual profile shapes are quite similar, although
our data provide hints of slightly higher pressures at both the smallest
and largest radii.
This agreement provides further evidence that the average cluster pressure
profile is approximately universal
(at least within our measurement uncertainties, which are $\simeq 10$--$20$\%
within $R_{500}$), especially given the large differences 
in sample masses and redshifts between the BOXSZ and these previous
analyses.
In addition, since many of the previous analyses relied on X-ray data, 
rather than SZ data, our results show the good agreement of
SZ and X-ray measurements of the ICM 
(\citealt{plagge10, melin11, planck11_x, komatsu11, bonamente11}, P12).

Finally, we simultaneously fit for the overall mass scaling,
the ensemble mean profile, and 
the radius-dependent intrinsic scatter of the 
pressure profiles using a Gaussian process model.
We find that the fractional scatter is minimized at radii between $\simeq 0.2R_{500}$
and $\simeq 0.5R_{500}$ at values $\lesssim 20$\%,
with larger scatter at both smaller and larger radii.
The best-fit mass scaling has a power law slope of 0.49
(compared to the nominal prediction of 2/3 based on self-similar hydrostatic
equilibrium models),
which is nearly identical to the scaling expected from the
X-ray derived $Y_X$--$M$ scaling determined by M10
using an identical X-ray mass determination.
Given our measurement uncertainties, 
our intrinsic scatter constraints as a function of radius are
consistent with previous analyses
which have largely relied on X-ray measurements of lower mass
and lower redshift clusters \citep[A10, ][P12]{sun11}.
This result provides additional evidence that pressure 
profiles are approximately universal over a wide range of masses
and redshifts, with intrinsic scatter of $\simeq 20$--$40$\% about
the universal profile, and that SZ and X-ray measurements of these profiles
are consistent with each other given current observational precision.

\section{Acknowledgments}

We acknowledge the assistance of: 
the day crew and Hilo
staff of the Caltech Submillimeter Observatory, who provided
invaluable assistance during data-taking for this
data set; 
Kathy Deniston, Barbara Wertz, and Diana Bisel, who provided effective
administrative support at Caltech and in Hilo;
Matt Hollister and Matt Ferry, who assisted in the
collection of these data;
the referee, who provided numerous useful suggestions.
The Bolocam observations were supported by the Gordon and Betty
Moore Foundation.
JS was supported by a NASA Graduate Student Research Fellowship,
a NASA Postdoctoral Program Fellowship, NSF/AST-0838261
and NASA/NNX11AB07G;
TM was supported by NASA through the Einstein Fellowship Program grant
PF0-110077;
NC was partially supported by a NASA Graduate Student
Research Fellowship;
AM was partially supported by NSF/AST-0838187;
SA, EP, and JAS were partially supported by NASA/NNX07AH59G;
KU acknowledges partial support from the National Science Council
of Tawain grant NSC100-2112-M-001-008-MY3 and from the Academia Sinica Career
Development Award.
A portion of this research was carried out at the Jet Propulsion
Laboratory, California Institute of Technology, under a contract
with the National Aeronautics and Space Administration.
This research made use of the Caltech Submillimeter Observatory,
which is operated by the California Institute of Technology
under cooperative agreement with the National Science Foundation 
(NSF/AST-0838261).

\end{document}